\documentclass{article}
\usepackage{pr3efCOQ}

\usepackage{hyperref}
\makeatletter
\hypersetup{
pdftitle={\@title},
pdfauthor={\@author},
breaklinks=true, 
colorlinks=true,
 linkcolor=indblue,
 citecolor=kwred,
 urlcolor=defgreen,
}
\makeatother

\begin{document}
      
\title{Formal verification in {Coq} of program properties involving
  the global state effect}
\author{Jean-Guillaume Dumas\thanks{LJK, Universit\'e de Grenoble, France.
     \href{mailto:Jean-Guillaume.Dumas@imag.fr,Dominique.Duval@imag.fr,Burak.Ekici@imag.fr}{\{Jean-Guillaume.Dumas,Dominique.Duval,Burak.Ekici\}@imag.fr}} 
   \and Dominique Duval\footnotemark[1] 
   \and  Burak Ekici\footnotemark[1]
   \and Damien Pous\thanks{LIP, ENS Lyon, France. \href{mailto:Damien.Pous@ens-lyon.fr}{Damien.Pous@ens-lyon.fr}}}
	\maketitle
	\begin{abstract}
The syntax of an imperative language does not mention explicitly the
state, while its denotational semantics has to mention it.
In this paper we present a framework for the verification in Coq of
properties of programs manipulating the global state effect.
These properties are expressed in a proof system which is close to the
syntax, as in effect systems, in the sense that the state does not appear
explicitly in the type of expressions which manipulate it. 
Rather, the state appears via decorations added to terms and to
equations. 
In this system, proofs of programs thus present two aspects: 
properties can be verified {\em up to effects} or the effects can be taken into
account. 
The design of our Coq library consequently reflects these two aspects: our
framework is centered around the construction of two inductive and dependent
types, one for terms up to effects and one for the manipulation of decorations.
	\end{abstract}  	
	\section{Introduction}
	\label{sec:intro}

The evolution of the state of the memory in an imperative program is a
computational effect: the state is never mentioned as an argument or a result of
a command, whereas in general it is used and modified during the execution of
commands. Thus, the syntax of an imperative language does not mention explicitly
the state, while its denotational semantics has to mention it. This means that
the state is encapsulated: its interface, which is made of the functions for
looking up and updating the values of the locations, is separated from its
implementation; the state cannot be accessed in any other way than through its 
interface.

It turns out that equational proofs in an imperative language may also
encapsulate the state: proofs can be performed without any knowledge of the
implementation of the state.  
This is made possible by adding decorations to terms, as in
effect systems~\cite{DBLP:conf/popl/LucassenG88,DBLP:journals/tocl/WadlerT03},
or by adding decorations to both the terms and the equations~\cite{DBLP:journals/corr/abs-1112-2396}.
The latter approach uses categorical constructions to model the
denotational semantics of the state effect and prove some properties of programs
involving this effect. \emph{Strong monads}, introduced by
Moggi~\cite{Moggi89notionsof}, were the first categorical approach to
computational effects, while Power et al.~\cite{Power:1997:PCN:967340.967345}
then proposed the \emph{premonoidal categories}. Next
Hughes~\cite{Hughes98generalisingmonads} extended Haskell with \emph{arrows} 
that share
some properties with the approach of \emph{cartesian effect categories} of Dumas
et al.~\cite{DBLP:journals/jsc/DumasDR11}.

The goal of this paper is to propose a Coq environment where proofs, written in
the latter decorated framework for the state effect, can be mechanised.

Proving properties of programs involving the state effect is important when the
order of evaluation of the arguments is not specified or more generally when
parallelization comes into play~\cite{DBLP:conf/popl/LucassenG88}. 
Indeed, pure computations, i.e. those not having any side-effects (or in other
words not modifying the state), are independent and could thus be run in
parallel. Differently, computations depending on or modifying the state should
be handled with more care. 

Now, proofs involving side-effects can become quite complex in order to be
fully rigorous. We will for instance look at the following property in details:
{\em recovering the value of a variable and setting up the value of another variable
can be performed in any order}. Such properties have been formalized for
instance by Plotkin et al.~\cite{DBLP:conf/fossacs/PlotkinP02} but the full
mathematical proof of such properties can be quite large.
The decorated approach of~\cite{DBLP:journals/corr/abs-1112-2396} helps since it
enables a verification of such proofs in two steps: a first step checks the
syntax {\em up to effects} by dropping the decorations; a second step then
takes the effects into account. 

To some extent, our work looks quite similar to \cite{Bertot:Kahn09} 
in the sense that we also define our own programming language 
and verify its properties by using axiomatic semantics. 
We construct our system on categorical notions (e.g. monads) 
as done in \cite{DBLP:journals/corr/abs-1012-1010}. In brief, we first declare our system components including their properties and then prove some related propositions. In that manner, the overall idea is also quite close 
to~\cite{stewart2013netcorewp}, even though technical details completely differ.

In this paper, we show that the decorated proof system can be developed
in Coq thus enabling a mechanised verification of decorated proofs for
side-effect systems. We recall in Section~\ref{sec:lefep} the logical
environment for decorated equational proofs involving the state effect. Then in
Section~\ref{sec:envcoq} we present the translation of the categorical rules
into Coq as well as their resulting derivations and the necessary
additions. The resulting Coq code has been integrated into a library, available
there: \url{http://coqeffects.forge.imag.fr}. Finally, in
Section~\ref{sec:cupdlkp} we give the full details of the proof of the property
above and its verification in Coq, as an example of the capabilities of our
library. Appendix~\ref{sec:decorules} is added for the sake of completeness
and readability in order to give the logical counterparts of the rules verified
in our Coq library.

\section{The Logical Environment for Equational Proofs}
\label{sec:lefep}

\subsection{Motivation}
\label{ssec:motiv}

Basically, in a purely functional programming language, 
an operation or a term $f$ 
with an argument of type $X$ and a result of type $Y$,
which may be written $f:X\to Y$ (in the \emph{syntax}), 
is interpreted (in the \emph{denotational semantics})
as a function $\deno{f}$ between the sets $\deno{X}$
and $\deno{Y}$, interpretations of $X$ and $Y$.
It follows that, when an operation has several arguments, 
these arguments can be evaluated in parallel, or in any order.
It is possible to interpret a purely functional
programming language via a categorical semantics based on
\emph{cartesian closed categories}; the word ``cartesian'' here
refers to the categorical \emph{products}, which are 
interpreted as \emph{cartesian products} of sets, 
and which are used for dealing with pairs (or tuples) of arguments. 
The \emph{logical semantics} of the language 
defines a set of rules that may be used for 
proving properties of programs.

But non-functional programming languages such as \texttt{C} or \texttt{Java} do
include computational effects. 
For instance a \texttt{C} function may modify the state structure and a
\texttt{Java} function may throw an exception during the computation. 
Such operations are examples of computational effects. 
In this paper we focus on the states effect.
We consider the $\lookup$ and $\update$ operations for
modeling the behavior of imperative programs: 
namely an $\update$ operation assigns a value to a location (or variable)  
and a $\lookup$ operation recovers the value of a location.
There are many ways to handle computational effects in programming languages. 
Here we focus on the categorical treatment
of~\cite{DBLP:journals/jsc/DumasDR11}, adapted to the state effect~\cite
{DBLP:journals/corr/abs-1112-2396}: 
this provides a logical semantics relying on {\em decorations}, or annotations,
of terms and equations.

\subsection{Decorated functions and equations for the states effect} 

\label{ssec:decorations}

The functions in our language 

are classified according to the way they interact with the state. 
The classification takes the form of annotations, or decorations, 
written as superscripts.
A function can be a \emph{modifier}, an \emph{accessor}
or a \emph{pure} function.
\begin{itemize}
\item As the name suggests, a \emph{modifier} may modify or use the state:
it is a \emph{read-write} function. We will use the keyword $\modi$ 
as an annotation for modifiers.
\item An \emph{accessor} may use the state structure but never modifies it:
it is a \emph{read-only} function. We will use the keyword $\acc$ 
for accessors. 
\item A \emph{pure function} never interacts with the state.  
We will use the keyword $\pure$ for pure functions.
\end{itemize}
The denotational semantics of this language is given in terms of 
the set of states $S$ and the 
\emph{cartesian product} operator \lq{}$\times$\rq{}. 
For all types $X$ and $Y$, interpreted as sets $\deno{X}$ and $\deno{Y}$, 
a modifier function $f:X\to Y$ is interpreted as 
a function $\deno{f} : \deno{X} \times S \to \deno{Y} \times S$ 
(it can access the state and modify it); 
an accessor $g$ as $\deno{g} : \deno{X} \times S \to \deno{Y}$ 
(it can access the state but not modify it); 
and a pure function $h$ as $\deno{h} : \deno{X} \to \deno{Y}$ 
(it can neither access nor modify the state).
There is a hierarchy among those functions. 
Indeed any pure function can be seen as both an accessor or a
modifier even though it will actually do not make use of its argument $S$. 
Similarly an accessor can be seen as a modifier.

The state is made of memory \emph{locations}, or \emph{variables};  
each location has a value which can be updated.  
For each location $i$, let $V_i$ be the type of the values 
that can be stored in the location $i$, and let $\Val_i=\deno{V_i}$ 
be the interpretation of $V_i$. 
In addition, the unit type is denoted by $\unit$; 
its interpretation is a singleton, it will also be denoted by $\unit$. 

The assignment of a value of type $V_i$ to a variable $i$  
takes an argument of type $V_i$.
It does not return any result but it modifies the state: 
given a value $a \in \Val_i$, 
the assignment of $a$ to $i$ sets the value of 
location $i$ to $a$ and keeps the value of the other 
locations unchanged.
Thus, this operation is a modifier from $V_i$ to $\unit$. 
It is denoted by $\update_i^{\modi}:V_i\to \unit$ 
and it is interpreted as $\deno{\update_i}:\Val_i \times S \to S$. 

The recovery of the value stored in a location $i$ 
takes no argument an returns a value of type $V_i$.
It does not modify the state but it observes the value stored 
at location $i$.
Thus, this operation is an accessor from $\unit$ to $V_i$.
It is denoted by $\lookup_i^{\acc}:\unit \to V_i$ 
and it is interpreted (since $\unit\times S$ is in bijection
with $S$) as $\deno{\lookup_i}:S \to \Val_i $. 

For each type $X$, 
the \emph{identity} operation $id_X: X \to X$, which is interpreted by 
mapping each element of $\deno{X}$ to itself, is pure.

Similarly, the \emph{final} operation $\tu_X: X \to \unit$, 
which is interpreted by 
mapping each element of $\deno{X}$ to the unique element 
of the singleton $\unit$, is pure. 
In order to lighten the notations we will often use $id_i$ and $\tu_i$ 
instead of respectively $id_{\Val_i}$ and $\tu_{\Val_i}$.

In addition, decorations are also added to equations.
\begin{itemize}
\item Two functions $f, g: X \to Y$ are \emph{strongly equal} if they 
return the same result and have the same effect on the state structure.
This is denoted $f \eqs g$. 
\item Two functions $f, g: X \to Y$ are \emph{weakly equal} 
if they return the same result but may have different effects on the state. 
This is denoted $f \eqw g$.
\end{itemize}

The state can be observed thanks to the lookup functions.
For each location $i$, the interpretation of the $\update_i$ operation
is characterized by the following equalities, 
for each state $s\in S$ and each $x\in \Val_i$: 
$$\begin{cases}
\deno{\lookup_i}(\deno{\update_i}(s,x)) = x   \\
\deno{\lookup_j}(\deno{\update_i}(s,x)) = \deno{\lookup_j}(s) 
  \mbox{ for every } j \in \Loc,\; j\ne i \\
\end{cases}$$
According to the previous definitions, 
these equalities are the interpretations of the following weak equations:
$$\begin{cases}
\lookup_i^{\acc} \circ \update_i^{\modi} \eqw id_i^{\pure} \colon V_i\to V_i \\
\lookup_j^{\acc} \circ \update_i^{\modi} \eqw 
  lookup_j^{\acc} \circ \tu_i^{\pure} \;
  \mbox{ for every } j \in \Loc,\; j\ne i  \colon V_i\to V_j \\ 
\end{cases}$$

\subsection{Sequential products}
\label{ssec:seqprod}

In functional programming, 
the product of functions allows to model operations with several arguments. 
But when side-effects occur (typically, updates of the state),
the result of evaluating the arguments 
may depend on the order in which they are evaluated. 
Therefore, we use \emph{sequential products} of functions, 
as introduced in \cite{DBLP:journals/jsc/DumasDR11}, 
which impose some order of evaluation of the arguments:
a sequential product is obtained as the sequential
composition of two \emph{semi-pure products}.
A semi-pure product, as far as we are concerned in this paper,
is a kind of product of an identity function (which is pure) 
with another function (which may be any modifier).

For each types $X$ and $Y$, we introduce a \emph{product} type $X\times Y$
with \emph{projections} $\pi_{1,X_1,X_2}^\pure: X_1 \times X_2 \to X_1$ and 
$\pi_{2,X_1,X_2}^\pure: X_1 \times X_2 \to X_2$,
which will be denoted simply by $\pi_1^\pure$ and $\pi_2^\pure$. 
This is interpreted as the cartesian product with its projections.
Pairs and products of pure functions are built as usual.
In the special case of a product with the unit type, 
it can easily be proved, as usual, that $\pi_1^\pure:X \times \unit\to X$ is
invertible with inverse the pair 
$(\pi_1^{-1})^\pure=\tuple{\id_X^\pure,\tu_X^\pure}\colon X\to X \times \unit$, 
and that $\pi_2^\pure=\tu_X^\pure:X \times \unit\to \unit$.
The \emph{permutation} operation $\perm_{X\times Y}: X \times Y \to Y \times X$
is also pure: it is interpreted as the function which exchanges
its two arguments. 

Given a pure function $f_1^{\pure}\colon X\to Y_1$, interpreted as  $\deno{f_1}:\deno{X}\to \deno{Y_1}$, and a modifier $f_2^{\modi}\colon X\to Y_2$  with its interpretation $\deno{f_2}:\deno{X}\times S\to \deno{Y_2}\times S$,
the \emph{left semi-pure pair} $\ltuple{f_1,f_2}^\modi\colon X\to Y_1\times Y_2$ 
is the modifier interpreted by $ \deno{\ltuple{f_1,f_2}} \colon 
\deno{X}\times S\to \deno{Y_1}\times \deno{Y_2}\times S $
such that  $ \deno{\ltuple{f_1,f_2}}(x,s) = 
(y_1,y_2,s')$ where $y_1 = \deno{f_1}(x)$ and $(y_2,s')=\deno{f_2}(x,s)$.  
The left semi-pure pair $\ltuple{f_1,f_2}^\modi$ is characterized, 
up to strong equations, by a weak and a strong equation:
  $$  \pi_1^\pure \circ \ltuple{f_1,f_2}^\modi \eqw f_1^{\pure} 
 \mbox{ and } \pi_2^\pure \circ \ltuple{f_1,f_2}^\modi \eqs f_2^{\modi} $$ 
The \emph{right semi-pure pair} $\rtuple{f_1,f_2}^\modi\colon X\to Y_1\times Y_2$ where $f_1^{\modi}: X \to Y_1$ and $f_2^{\pure}: X \to Y_2$
is defined in the symmetric way:

$$
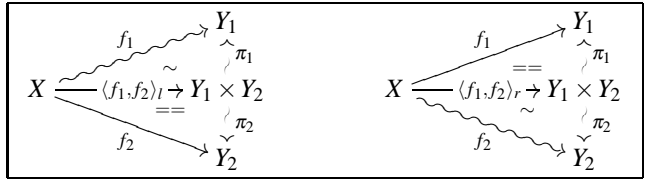
\begin{array}{|ccc|}
\hline
\xymatrix@R=1pc@C=4pc{
& Y_1 \\ 
X \ar@{~>}[ru]^{f_1} \ar[rd]_{f_2} 
  \ar[r]|{\;\ltuple{f_1,f_2}\;} & 
  Y_1\times Y_2  \ar@{}[lu]|(0.3){\eqw} \ar@{}[ld]|(0.3){\eqs}
     \ar@{~>}[u]_{\pi_1} \ar@{~>}[d]^{\pi_2} \\
& Y_2 \\ 
} &
\qquad &
\xymatrix@R=1pc@C=4pc{
& Y_1 \\ 
X \ar@{~>}[rd]_{f_2} \ar[ru]^{f_1} 
  \ar[r]|{\;\rtuple{f_1,f_2}\;} & 
  Y_1\times Y_2   \ar@{}[lu]|(0.3){\eqs} \ar@{}[ld]|(0.3){\eqw}
     \ar@{~>}[u]_{\pi_1} \ar@{~>}[d]^{\pi_2} \\
& Y_2 \\ 
} \\ 
\hline
\end{array}
$$

{\bf Note.} In all diagrams, the decorations are expressed by shapes 
of arrows: 
waving arrows for pure functions, 
dotted arrows for accessors and straight arrows for modifiers.

The \emph{left semi-pure product} 
is defined in the usual way from the left semi-pure pair: 
given $f_1^\pure\colon X_1\to Y_1$ $f_2^\modi\colon X_2\to Y_2$, the left semi-pure product
of $f_1$ and $f_2$ is $(f_1\ltimes f_2)^\modi
= \ltuple{f_1 \circ \pi_{1,X_1,X_2}, f_2\circ \pi_{2,X_1,X_2}} ^\modi 
\colon X_1\times X_2\to Y_1\times Y_2$. 
It is characterized, 
up to strong equations, by a weak and a strong equation:
  $$  \pi_{1,Y_1,Y_2}^\pure \circ (f_1\ltimes f_2)^\modi \eqw f_1^\pure \circ \pi_{1,X_1,X_2}^\pure 
 \mbox{ and } \pi_{2,Y_1,Y_2}^\pure \circ (f_1\ltimes f_2)^\modi \eqs 
   f_2^\modi \circ \pi_{2,X_1,X_2}^\pure $$

The \emph{right semi-pure product} $(f_1 \rtimes f_2)^\modi
\colon X_1\times X_2\to Y_1\times Y_2$ 
is defined in the symmetric way:

$$
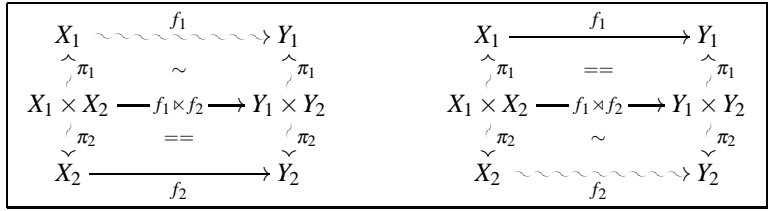
\begin{array}{|ccc|}
\hline
\xymatrix@R=1pc@C=4pc{
X_1 \ar@{~>}[r]^{f_1} & Y_1 \\ 
X_1\times X_2 
  \ar[r]|{\;f_1\ltimes f_2\;} 
     \ar@{~>}[u]_{\pi_1} \ar@{~>}[d]^{\pi_2} &
  Y_1\times Y_2  \ar@{}[lu]|{\eqw} \ar@{}[ld]|{\eqs}
     \ar@{~>}[u]_{\pi_1} \ar@{~>}[d]^{\pi_2} \\
X_2 \ar[r]_{f_2} & Y_2 \\ 
} &
\qquad &
\xymatrix@R=1pc@C=4pc{
X_1 \ar[r]^{f_1} & Y_1 \\ 
X_1\times X_2 
  \ar[r]|{\;f_1 \rtimes f_2\;} 
     \ar@{~>}[u]_{\pi_1} \ar@{~>}[d]^{\pi_2} &
  Y_1\times Y_2  \ar@{}[lu]|{\eqs} \ar@{}[ld]|{\eqw}
     \ar@{~>}[u]_{\pi_1} \ar@{~>}[d]^{\pi_2} \\
X_2 \ar@{~>}[r]_{f_2} & Y_2 \\ 
} \\ 
\hline
\end{array}
$$

Now, it is easy to define the \emph{left sequential product} 
of two modifiers $f_1^\modi:X_1\to Y_1$ and $f_2^\modi:X_2\to Y_2$ 
by composing a right semi-pure product with a left semi-pure one and using $id$ function as the pure component:
  $$ (f_1 \ltimes f_2)^\modi = (\id_{Y_1} \ltimes f_2)^\modi
\circ (f_1 \rtimes \id_{X_2})^\modi \colon X_1\times X_2 \to Y_1\times Y_2 $$ 
In a symmetric way, the \emph{right sequential product}
of $f_1^\modi:X_1\to Y_1$ and $f_2^\modi:X_2\to Y_2$ is 
defined as:
   $$ (f_1 \rtimes f_2)^\modi = 
(f_1 \rtimes \id_{Y_2})^\modi \circ (\id_{X_1} \ltimes f_2)^\modi
\colon X_1\times X_2 \to Y_1\times Y_2 $$ 
The left sequential product models the fact of executing $f_1$ before $f_2$,
while the right sequential product models the fact of executing 
$f_2$ before $f_1$; in general they return different results
and they modify the state in a different way.

$$
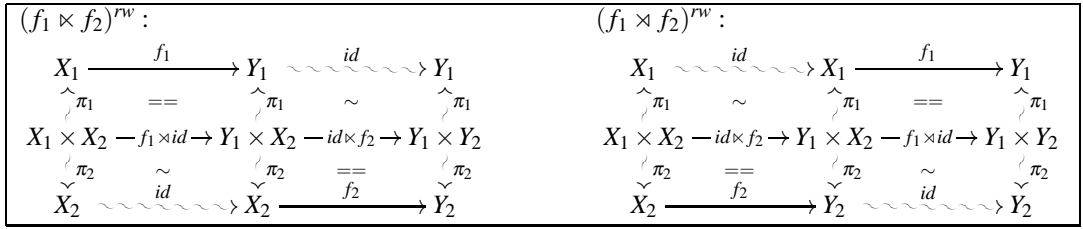
\begin{array}{|lll|}
\hline
(f_1 \ltimes f_2)^\modi: && 
(f_1 \rtimes f_2)^\modi: \\
\xymatrix@R=1pc@C=3pc{
X_1 \ar[r]^{f_1} & 
  Y_1 \ar@{~>}[r]^{\id} & 
  Y_1 \\ 
X_1\times X_2 \ar[r]|{\;f_1 \rtimes \id\;}  
     \ar@{~>}[u]_{\pi_1} \ar@{~>}[d]^{\pi_2} &
  Y_1\times X_2 \ar[r]|{\;\id\ltimes f_2\;}  
     \ar@{}[lu]|{\eqs} \ar@{}[ld]|{\eqw}
     \ar@{~>}[u]_{\pi_1} \ar@{~>}[d]^{\pi_2} &
  Y_1\times Y_2 
     \ar@{}[lu]|{\eqw} \ar@{}[ld]|{\eqs}
     \ar@{~>}[u]_{\pi_1} \ar@{~>}[d]^{\pi_2} \\ 
X_2 \ar@{~>}[r]^{\id} & 
  X_2 \ar[r]^{f_2} & 
  Y_2 \\ 
} &
\qquad &
\xymatrix@R=1pc@C=3pc{
X_1 \ar@{~>}[r]^{\id} & 
  X_1 \ar[r]^{f_1} & 
  Y_1 \\ 
X_1\times X_2 \ar[r]|{\;\id\ltimes f_2\;}  
     \ar@{~>}[u]_{\pi_1} \ar@{~>}[d]^{\pi_2} &
  Y_1\times X_2 \ar[r]|{\;f_1 \rtimes \id\;}
     \ar@{}[lu]|{\eqw} \ar@{}[ld]|{\eqs}
     \ar@{~>}[u]_{\pi_1} \ar@{~>}[d]^{\pi_2} &
  Y_1\times Y_2 
     \ar@{}[lu]|{\eqs} \ar@{}[ld]|{\eqw}
     \ar@{~>}[u]_{\pi_1} \ar@{~>}[d]^{\pi_2} \\ 
X_2 \ar[r]^{f_2} & 
  Y_2 \ar@{~>}[r]^{\id} & 
  Y_2 \\ 
} \\ 
\hline
\end{array}
$$

\subsection{A property of states}
\label{ssec:prop}

In \cite{DBLP:conf/fossacs/PlotkinP02} an equational presentation of states is
given, with seven equations. 
These equations are expressed as decorated equations 
in~\cite{DBLP:journals/corr/abs-1112-2396}.
They are the archetype of the properties of the proofs we want to
verify. 
For instance, the fact that modifying a location $i$ 
and observing the value of another location $j$ can be done in any order 
is called the \emph{commutation update-lookup} property.  
This property can be expressed as an equation relating the 
functions $\deno{\update_i}$ and $\deno{\lookup_j}$. 
For this purpose, let $\deno{\lookup_j}':S\times \Val_j\times S$ 
be defined by 
 $$ \deno{\lookup_j}'(s) = (s, \deno{\lookup_j}(s)) 
  \mbox{ for each } s\in S \;.$$
Thus, given a state $s$ and a value $a\in\Val_i$, 
assigning $a$ to $i$ and then observing the value of $j$ 
is performed by the function:
  $$ \deno{\lookup_j}' \circ\deno{\update_i} \colon 
  \Val_i \times S \to \Val_j \times S \;.$$
Observing the value of $j$ and then assigning $a$ to $i$ 
also corresponds to a function from 
$\Val_i \times S$  to $\Val_j \times S $ built from 
$\deno{\update_i}$ and $\deno{\lookup_j}'$. 
This function first performs $\deno{\lookup_j}'(s)$ while keeping $a$ unchanged,
then it performs $\deno{\update_i}(s,a)$ while keeping $b$ unchanged 
(where $b$ denotes the value of $j$ in $s$
which has been returned by $\deno{\lookup_j}(s)$).
The first step is $ \id_{\Val_i} \times \deno{\lookup_j}' \colon
\Val_i \times S \to \Val_i \times (\Val_j \times S)$ 
and the second step is 
$ \id_{\Val_j} \times \deno{\update_i} \colon
\Val_j \times (\Val_i \times S) \to \Val_j \times S$.
An intermediate permutation step is required, it is called 
$ \perm_{i,j} \colon \Val_i \times (\Val_j \times S) \to 
\Val_j \times (\Val_i \times S)$ such that $\perm_{i,j}(a,(b,s))=(b,(a,s))$.

Altogether, observing the value of $j$ and then assigning $a$ to $i$ 
corresponds to the function:
 $$ (\id_{\Val_j} \times \deno{\update_i}) 
\circ \perm_{i,j}  \circ 
(\id_{\Val_i} \times \deno{\lookup_j}') \colon
\Val_i \times S \to \Val_j \times S $$ 
Thus, the commutation update-lookup property means that: 
$$ \deno{\lookup_j}' \circ\deno{\update_i} = 
(\id_{\Val_j} \times \deno{\update_i}) 
\circ \perm_{i,j}  \circ 
(\id_{\Val_i} \times \deno{\lookup_j}') $$ 

According to Section~\ref{ssec:decorations}, 
this is the interpretation of the following strong equation,
which can also be expressed as a diagram:

\begin{equation}
\label{eq:cupdlkp}
 \lookup_j^\acc \circ update_i^\modi \eqs 
  \pi_2^\pure \circ (update_i^\modi \rtimes \id_j^\pure) 
  \circ (id_i^\pure \ltimes lookup_j^\acc) 
  \circ (\pi_1^{-1})^\pure :  V_i \to V_j \;.
\end{equation}

$$
\begin{array}{|ccc|}
\hline
\xymatrix@R=2pc@C=3pc{
V_i \ar[r]^{update_i} & \unit  \ar@{.>}[r]^{lookup_j} & V_j \\ 
} &
\eqs &
\xymatrix@R=2pc@C=3pc{
V_i \ar@{~>}[r]^{\pi_1^{-1}} & 
V_i \times \unit \ar@{.>}[r]^{\id_i \ltimessp lookup_j} & 
V_i\times V_j \ar[r]^{update_i \rtimessp \id_j} & 
\unit\times V_j \ar@{~>}[r]^{\pi_2} &
V_j \\
} \\ 
\hline
\end{array}
$$

\textbf{Remark.} Using the right sequential product, 
the right hand-side of the 
commutation update-lookup equation can be written as 
$\pi_2^\pure \circ (update_i^\modi \rtimes lookup_j^\acc) 
  \circ (\pi_1^{-1})^\pure $.
In addition, using the left sequential product, 
it is easy to check that 
the left hand-side of this equation can be written as 
$\pi_2^\pure \circ (update_i^\modi \ltimes lookup_j^\acc) 
  \circ (\pi_1^{-1})^\pure $.
Since $\pi_1^\pure:V_i\times\unit\to V_i$ and 
$\pi_2^\pure:\unit\times V_j\to V_j$  are invertible, 
we get a symmetric expression for the equation
which corresponds nicely to the description of the 
commutation update-lookup property as 
``the fact that modifying a location $i$ 
and observing the value of another location $j$ 
can be done in any order'': 
  $$  update_i^\modi \rtimes lookup_j^\acc \eqs 
  update_i^\modi \ltimes lookup_j^\acc $$

\section{The Environment in Coq}
\label{sec:envcoq}

In this Section we present the core of this paper, 
namely the implementation in the Coq proof assistant 
of the rules for reasoning with decorated operations and equations 
and the proof of the update-lookup commutation property 
using these rules.

In the preceding section, we have shown proofs of propositions 
involving effects.
We now present the construction of a Coq framework enabling one 
to formalize such
proofs.
This framework has been released as \texttt{STATES-0.5} library and is available 
in the following
web-site: \url{http://coqeffects.forge.imag.fr}.

In order to construct this framework, we need to define data structures, terms,
decorations and basic rules as axioms. Those give rise to derived rules and
finally to proofs. This organization is reflected in the library with
corresponding Coq modules, as shown 
in the following diagram:

	$$ \begin{array}{| c |} 
		\hline
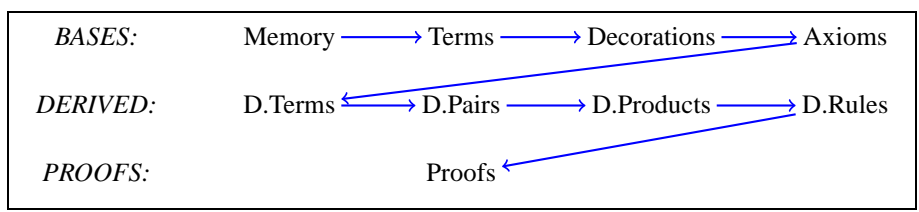
		\begin{dependency}
			\begin{deptext}[column sep=1cm, row sep=3ex]
				\emph{BASES:} \& Memory \& Terms \& Decorations \& Axioms \\ 
				\emph{DERIVED:} \& D.Terms \& D.Pairs \& D.Products \& D.Rules \\ 
				\emph{PROOFS:} \& \& Proofs \\
			\end{deptext}
			\draw [->, thick, blue] (\wordref{1}{2})--(\wordref{1}{3});
			\draw [->, thick, blue] (\wordref{1}{3})--(\wordref{1}{4});
			\draw [->, thick, blue] (\wordref{1}{4})--(\wordref{1}{5});
			\draw [->, thick, blue] (\wordref{1}{5})--(\wordref{2}{2});
			\draw [->, thick, blue] (\wordref{2}{2})--(\wordref{2}{3});
			\draw [->, thick, blue] (\wordref{2}{3})--(\wordref{2}{4});
			\draw [->, thick, blue] (\wordref{2}{4})--(\wordref{2}{5});
			\draw [->, thick, blue] (\wordref{2}{5})--(\wordref{3}{3});
		\end{dependency} \\
	\hline
 	\end{array}$$

The memory module uses declarations of \emph{locations}. 
A \emph{location} represents a field on the memory to store and observe data. 
Then terms are defined in steps. First we give the definitions of
\emph{non-decorated terms}: they constitute the main part of the design with the
inclusion of all the required functions. 
For instance, the \texttt{lookup} function which observes the current state is
defined from void ($\unit$, the terminal object of the underlying category) to
the set of values that could be stored in that specified location. 

The next step is to decorate those functions with respect to their manipulation
abilities on the state structure. 
For instance, the \texttt{update} function is defined as a \emph{modifier}. the
{\em modifier} status is represented by a \emph{rw} label in the library. 
All the rules related to decorated functions are stated in the module called
\emph{Axioms}.

Then, based on the ones already defined, some other terms are derived. 
For example, the derived \texttt{permut} function takes projections as the basis
and replaces the orders of input objects in a \emph{categorical product}. 
Similarly, by using the already defined rules (given in the \emph{axioms}
section), some additional rules are derived concerning \emph{categorical pairs},
\emph{products} and \emph{others} pointing the rules constructed over the ones
from different sources.

In the following subsection we detail the system sub-modules. The order of
enumeration gives the dependency among sub-modules as shown in 
the above diagram.

For instance, the module \emph{decorations} requires
definitions from the \emph{memory} and the \emph{terms} modules.
Then, as an example, we give the full proof, in Coq, of the
update-lookup commutation property of~\cite{DBLP:conf/fossacs/PlotkinP02}.

		\subsection{Proof System for States}
		\label{sec:psfstates}
In this section we give the Coq definitions of our proof system and explore them
module by module.

The major ideas in the construction of this Coq framework are:
\begin{itemize}
\item All the features of the proof system, that are given in the previous Sections~\ref{ssec:decorations}, \ref{ssec:seqprod}, \ref{ssec:prop} and in the appendix \ref{sec:decorules}, definitely constitute the basis for the Coq implementation. In brief, we first declare all the terms without decorations, then we decorate them and after all we end up with the rules involving decorated terms. We also confirm that if one removes all the decorations (hence transforming every operations into \coqdocconstructor{pure} terms), the proof system remains valid.
\item The terms \texttt{pair} and \texttt{perm\_pair} express the construction of pairs of two functions. As shown in Section~\ref{ssec:seqprod}, in the presence of effects the order of evaluation matters. Therefore, we first define the left pair in Coq and simply call it \texttt{pair} (see Section~\ref{sec:dndecterms}). Then, the right pair can be {\em derived} using the permutation rule and it is called \texttt{perm\_pair} (see Section~\ref{sec:ddwdec}). 
\item The most challenging part of the design is the proof implementations of the propositions by \cite{DBLP:conf/fossacs/PlotkinP02}, since they are quite tricky and long. We assert implementations tricky, because to see the main schema (or flowchart) of the proofs at first sight and coding them in Coq with this reasoning is quite difficult. To do so, we first sketch the related diagrams with marked equalities (\emph{strong} or \emph{weak}), then we convert them into some line equations, representing the main propositions to be shown.
In order to do so, we use a fractional notation together with the exploited rules for each step. Eventually, Coq implementations are done by coding each step which took part in the fractional notation. From this aspect, without the fractional correspondences, proofs might be seen a little tough to follow. In order to  increase the readability score, we divided those implementations into sub-steps and gave the associated relevant explanations. See Section~\ref{sec:cupdlkp} for an example.
\item Considering the entire design, we benefit from an important aspect provided by Coq environment, namely \emph{dependent types}. They provide a unified formalism in the creation of new data types and allow us to deal in a simple manner with most of the typing issues. More precisely, the type \coqref{doc.term}{\coqdocinductive{term}} is not a \coqdockw{Type}, but rather a  
\coqdockw{Type} \ensuremath{\rightarrow} \coqdockw{Type} \ensuremath{\rightarrow} \coqdockw{Type}.
The domain/codomain information of \coqref{doc.term}{\coqdocinductive{term}} is embedded into Coq type system, so that we do not need to talk about ill-typed terms. For instance, \texttt{pi1 $\circ$ final} is ill-typed since \coqdocconstructor{final} is defined from any object \coqdocvar{X}: \coqdockw{Type} to  \coqexternalref{unit}{http://coq.inria.fr/distrib/8.3pl4/stdlib/Coq.Init.Datatypes}{\coqdocinductive{unit}} where \coqdocconstructor{pi1} if from \coqdocvar{Y}: \coqdockw{Type} to \coqdocvar{Z}: \coqdockw{Type}. Therefore, the latter composition cannot be seen as a \coqref{doc.term}{\coqdocinductive{term}}.
\end{itemize}
 		\subsection{Memory}
		\label{sec:prereq}
We represent the set of memory locations by a Coq parameter $\mathit{Loc:
  Type}$. Since memory locations may contain different types of values, we also
assume a function $\mathit Val: Loc \rightarrow Type$ that indicates the type of
values contained in each location.
		\subsection{Terms}
		\label{sec:dndecterms}
Non-decorated operators, using the monadic equational logic and categorical
products, are represented by an inductive Coq data type named
\coqdocinductive{term}.
It basically gets two Coq types, that are corresponding either to
objects or to mappings in the given categorical structure, and returns a
function type. 
Those function types are the representations of the homomorphisms of the
category.  
We summarize these non-decorated constructions below:
\coqdocemptyline
\coqdocindent{3.0em}
\coqdockw{Inductive} \coqdef{doc.term}{term}{\coqdocinductive{term}}: \coqdockw{Type} \ensuremath{\rightarrow} \coqdockw{Type} \ensuremath{\rightarrow} \coqdockw{Type} := \coqdoceol
\coqdocindent{5.00em}
\ensuremath{|} \coqdef{doc.id}{id}{\coqdocconstructor{id}}: \coqdockw{\ensuremath{\forall}} \{\coqdocvar{X}: \coqdockw{Type}\}, \coqref{Assumptions.term}{\coqdocinductive{term}} \coqdocvar{X} \coqdocvar{X}\coqdoceol
\coqdocindent{5.00em}
\ensuremath{|} \coqdef{doc.comp}{comp}{\coqdocconstructor{comp}}: \coqdockw{\ensuremath{\forall}} \{\coqdocvar{X} \coqdocvar{Y} \coqdocvar{Z}: \coqdockw{Type}\}, \coqref{Assumptions.term}{\coqdocinductive{term}} \coqdocvar{X} \coqdocvar{Y} \ensuremath{\rightarrow} \coqref{Assumptions.term}{\coqdocinductive{term}} \coqdocvar{Y} \coqdocvar{Z} \ensuremath{\rightarrow} \coqref{Assumptions.term}{\coqdocinductive{term}} \coqdocvar{X} \coqdocvar{Z} \coqdoceol
\coqdocindent{5.00em}
\ensuremath{|} \coqdef{doc.final}{final}{\coqdocconstructor{final}}: \coqdockw{\ensuremath{\forall}} \{\coqdocvar{X}: \coqdockw{Type}\}, \coqref{Assumptions.term}{\coqdocinductive{term}} \coqexternalref{unit}{http://coq.inria.fr/distrib/8.3pl4/stdlib/Coq.Init.Datatypes}{\coqdocinductive{unit}} \coqdocvar{X} \coqdoceol
\coqdocindent{5.00em}
\ensuremath{|} \coqdef{doc.pair}{pair}{\coqdocconstructor{pair}}: \coqdockw{\ensuremath{\forall}} \{\coqdocvar{X} \coqdocvar{Y} \coqdocvar{Z}: \coqdockw{Type}\}, \coqref{Assumptions.term}{\coqdocinductive{term}} \coqdocvar{X} \coqdocvar{Z} \ensuremath{\rightarrow} \coqref{Assumptions.term}{\coqdocinductive{term}} \coqdocvar{Y} \coqdocvar{Z} \ensuremath{\rightarrow} \coqref{Assumptions.term}{\coqdocinductive{term}} (\coqdocvar{X}\coqexternalref{:type scope:x '*' x}{http://coq.inria.fr/distrib/8.3pl4/stdlib/Coq.Init.Datatypes}{\coqdocnotation{\ensuremath{\times}}}\coqdocvar{Y}) \coqdocvar{Z} \coqdoceol
\coqdocindent{5.00em}
\ensuremath{|} \coqdef{doc.pi1}{pi1}{\coqdocconstructor{pi1}}: \coqdockw{\ensuremath{\forall}} \{\coqdocvar{X} \coqdocvar{Y}: \coqdockw{Type}\}, \coqref{Assumptions.term}{\coqdocinductive{term}} \coqdocvar{X} (\coqdocvar{X}\coqexternalref{:type scope:x '*' x}{http://coq.inria.fr/distrib/8.3pl4/stdlib/Coq.Init.Datatypes}{\coqdocnotation{\ensuremath{\times}}}\coqdocvar{Y}) \coqdoceol
\coqdocindent{5.00em}
\ensuremath{|} \coqdef{doc.pi2}{pi2}{\coqdocconstructor{pi2}}: \coqdockw{\ensuremath{\forall}} \{\coqdocvar{X} \coqdocvar{Y}: \coqdockw{Type}\}, \coqref{Assumptions.term}{\coqdocinductive{term}} \coqdocvar{Y} (\coqdocvar{X}\coqexternalref{:type scope:x '*' x}{http://coq.inria.fr/distrib/8.3pl4/stdlib/Coq.Init.Datatypes}{\coqdocnotation{\ensuremath{\times}}}\coqdocvar{Y}) \coqdoceol
\coqdocindent{5.00em}
\ensuremath{|} \coqdef{doc.lookup}{lookup}{\coqdocconstructor{lookup}}: \coqdockw{\ensuremath{\forall}} \coqdocvar{i}: \coqdocaxiom{Loc}, \coqref{Assumptions.term}{\coqdocinductive{term}} (\coqref{Assumptions.M.Val}{\coqdocaxiom{Val}} \coqdocvar{i}) \coqexternalref{unit}{http://coq.inria.fr/distrib/8.3pl4/stdlib/Coq.Init.Datatypes}{\coqdocinductive{unit}}	\coqdoceol
\coqdocindent{5.00em}
\ensuremath{|} \coqdef{doc.update}{update}{\coqdocconstructor{update}}: \coqdockw{\ensuremath{\forall}} \coqdocvar{i}: \coqdocaxiom{Loc}, \coqref{Assumptions.term}{\coqdocinductive{term}} \coqexternalref{unit}{http://coq.inria.fr/distrib/8.3pl4/stdlib/Coq.Init.Datatypes}{\coqdocinductive{unit}} (\coqref{Assumptions.M.Val}{\coqdocaxiom{Val}} \coqdocvar{i}).\coqdoceol
\coqdocemptyline
\coqdocindent{3.0em}
\coqdocvar{Infix }\coqdef{doc.::x 'o' x}{"}{\coqdocnotation{"}}o{\coqdocnotation{"}} := \coqref{doc.comp}{\coqdocconstructor{comp}} (\coqdoctac{at} \coqdocvar{level} 70).\coqdoceol
\coqdocemptyline

Note that a term of type \coqdocinductive{term} \coqdocvar{X Y} is interpreted as a function from the set
 \coqdocvar{Y} to the set  \coqdocvar{X} (the co-domain, \coqdocvar{X}, is given first.)

The constructor \coqdocconstructor{id} denotes the identity function: for any
type  \coqdocvar{X}, \coqdocconstructor{id} \coqdocvar{X} has type \coqdocinductive{term} \coqdocvar{X X}.
The term \coqdocconstructor{comp} composes two given compatible function types
and returns another one. 
The term \coqdocconstructor{pair} represents the categorical product type of two
given objects. For instance, if  \coqdocinductive{term} \coqdocvar{X Z}
corresponds to a mapping defined from an object  \coqdocvar{Z} to another one
denoted as  \coqdocvar{X}, then \coqdocconstructor{pair} with input types
\coqdocinductive{term} \coqdocvar{X Z} and \coqdocinductive{term} \coqdocvar{Y
  Z}, agreeing on domains, returns a new function type of form
\coqdocinductive{term} \coqdocvar{(X $\times$ Y) Z}. 
The terms \coqdocconstructor{pi1} and \coqdocconstructor{pi2} are projections of products while \coqdocconstructor{final} maps any object to
the terminal object (the singleton set, denoted by $\unit$) of the Cartesian
effect category in question.
\coqdocconstructor{lookup} takes a location
identifier and denotes the lookup operation for the relevant location. It is
mathematically defined from the terminal object of the category. 
As the name suggests, the \coqdocconstructor{update} operator updates the value
in the specified location. 
		\subsection{Decorations}
		\label{sec:dterms}

In order to keep the semantics of state close to syntax, all the operations are
decorated with respect to their manipulation abilities on the state structure.
In Coq, we define another inductive data type, called \coqdocinductive{kind}, to
represent these decorations.
Its constructors are \coqdocconstructor{pure} (decorated by $\pure$),
\coqdocconstructor{ro} (for read-only and decoration $\acc$) and
\coqdocconstructor{rw} (for read-write and decoration $\modi$). 
It should be recalled that if a function is \coqdocconstructor{pure}, then it
could be seen both as \coqdocconstructor{ro} (accessor) and
\coqdocconstructor{rw} (modifier), due to the hierarchy rule among decorated functions:
\coqdocemptyline
\coqdocindent{3.0em}
\coqdockw{Inductive} \coqdef{doc.kind}{kind}{\coqdocinductive{kind}} := \coqdef{doc.pure}{pure}{\coqdocconstructor{pure}} \ensuremath{|} \coqdef{doc.ro}{ro}{\coqdocconstructor{ro}} \ensuremath{|} \coqdef{doc.rw}{rw}{\coqdocconstructor{rw}}.
\coqdocindent{0.00em}
\coqdocemptyline
In Coq, we had to define the decorations of terms via the separate inductive
data type called \coqdocinductive{is}.
The latter takes a term and a kind and returns a \coqdockw{Prop}. 
In other words, \coqdocinductive{is} indicates whether the given term is allowed to
be decorated by the given kind or not. 
For instance, the term \coqref{doc.id}{\coqdocconstructor{id}} is
\coqref{doc.pure}{\coqdocconstructor{pure}}, since it cannot use nor modify the
state. 
Therefore it is by definition decorated with the keyword $\pure$.
This decoration is checked by a constructor \coqdocconstructor{is\_id}. 
To illustrate this, if one (by using \coqdoctac{apply} tactic of Coq) asks
whether \coqdocconstructor{id} is pure, then the returned result would be have
to be \coqdockw{True}. In order to check whether \coqdocconstructor{id} is an
accessor or a modifier, the constructors \coqdocconstructor{is\_pure\_ro} and
\coqdocconstructor{is\_ro\_rw} should be applied beforehand to convert both
statements into \coqdocinductive{is} \coqdocconstructor{pure id}.
The incidence of decorations upon the terms is summarized below together with
their related rules (detailed in Appendix~\ref{sec:decorules}):
\coqdocemptyline
\coqdocindent{-7pt}
\begin{tabular}{lcr}
\coqdockw{Inductive} \coqdef{doc.is}{is}{\coqdocinductive{is}}:
\coqref{doc.kind}{\coqdocinductive{kind}} \ensuremath{\rightarrow}
\coqdockw{\ensuremath{\forall}} \coqdocvar{X} \coqdocvar{Y},
\coqref{doc.term}{\coqdocinductive{term}} \coqdocvar{X} \coqdocvar{Y}
\ensuremath{\rightarrow} \coqdockw{Prop} := 
& \emph{Rule} & \emph{Fig.} \coqdoceol \\
\coqdocindent{5pt}
\ensuremath{|} \coqdef{doc.is id}{is\_id}{\coqdocconstructor{is\_id}}:
\coqdockw{\ensuremath{\forall}} \coqdocvar{X},
\coqref{Assumptions.is}{\coqdocinductive{is}}
\coqref{doc.pure}{\coqdocconstructor{pure}}
(@\coqref{doc.id}{\coqdocconstructor{id}} \coqdocvar{X}) 
& (0-id) & (\ref{fig:meqn}) \coqdoceol \\
\coqdocindent{5pt}
\ensuremath{|} \coqdef{doc.is comp}{is\_comp}{\coqdocconstructor{is\_comp}}:
\coqdockw{\ensuremath{\forall}} \coqdocvar{k} \coqdocvar{X} \coqdocvar{Y}
\coqdocvar{Z} (\coqdocvar{f}: \coqref{doc.term}{\coqdocinductive{term}}
\coqdocvar{X} \coqdocvar{Y}) (\coqdocvar{g}:
\coqref{doc.term}{\coqdocinductive{term}} \coqdocvar{Y} \coqdocvar{Z}),
\coqref{Assumptions.is}{\coqdocinductive{is}} \coqdocvar{k} \coqdocvar{f}
\ensuremath{\rightarrow} \coqref{Assumptions.is}{\coqdocinductive{is}}
\coqdocvar{k} \coqdocvar{g} \ensuremath{\rightarrow}
\coqref{Assumptions.is}{\coqdocinductive{is}} \coqdocvar{k} (\coqdocvar{f}
\coqref{doc.::x 'o' x}{\coqdocnotation{o}} \coqdocvar{g}) 
& (dec-comp)  & (\ref{fig:meqn}) \coqdoceol\\
\coqdocindent{5pt}
\ensuremath{|} \coqdef{doc.is final}{is\_final}{\coqdocconstructor{is\_final}}:
\coqdockw{\ensuremath{\forall}} \coqdocvar{X},
\coqref{Assumptions.is}{\coqdocinductive{is}}
\coqref{doc.pure}{\coqdocconstructor{pure}} (@\coqref{doc.final}{\coqdocconstructor{final}} \coqdocvar{X}) 
& (0-final)  & (\ref{fig:empty-prod-existence}) \coqdoceol \\
\coqdocindent{5pt}
\ensuremath{|} \coqdef{doc.is pair}{is\_pair}{\coqdocconstructor{is\_pair}}:
\coqdockw{\ensuremath{\forall}} \coqdocvar{k} \coqdocvar{X} \coqdocvar{Y}
\coqdocvar{Z} (\coqdocvar{f}: \coqref{doc.term}{\coqdocinductive{term}}
\coqdocvar{X} \coqdocvar{Z}) (\coqdocvar{g}:
\coqref{doc.term}{\coqdocinductive{term}} \coqdocvar{Y} \coqdocvar{Z}),
\coqref{Assumptions.is}{\coqdocinductive{is}} \coqdocvar{k} \coqdocvar{f}
\ensuremath{\rightarrow} \coqref{Assumptions.is}{\coqdocinductive{is}}
\coqdocvar{k} \coqdocvar{g} \ensuremath{\rightarrow}
\coqref{Assumptions.is}{\coqdocinductive{is}} \coqdocvar{k}
(\coqref{doc.pair}{\coqdocconstructor{pair}} \coqdocvar{f} \coqdocvar{g}) 
& (dec-pair-exists) &(\ref{fig:dec_lpair_unicity_existence}) \coqdoceol\\
\coqdocindent{5pt}
\ensuremath{|} \coqdef{doc.is pi1}{is\_pi1}{\coqdocconstructor{is\_pi1}}:
\coqdockw{\ensuremath{\forall}} \coqdocvar{X} \coqdocvar{Y},
\coqref{Assumptions.is}{\coqdocinductive{is}}
\coqref{doc.pure}{\coqdocconstructor{pure}}
(@\coqref{doc.pi1}{\coqdocconstructor{pi1}} \coqdocvar{X} \coqdocvar{Y})
& (0-proj-1) & (\ref{fig:bin-prod-existence}) \coqdoceol\\
\coqdocindent{5pt}
\ensuremath{|} \coqdef{doc.is pi2}{is\_pi2}{\coqdocconstructor{is\_pi2}}:
\coqdockw{\ensuremath{\forall}} \coqdocvar{X} \coqdocvar{Y},
\coqref{Assumptions.is}{\coqdocinductive{is}}
\coqref{doc.pure}{\coqdocconstructor{pure}}
(@\coqref{doc.pi2}{\coqdocconstructor{pi2}} \coqdocvar{X} \coqdocvar{Y}) 
& (0-proj-2)   & (\ref{fig:bin-prod-existence}) \coqdoceol\\
\coqdocindent{5pt}
\ensuremath{|} \coqdef{doc.is
  lookup}{is\_lookup}{\coqdocconstructor{is\_lookup}}:
\coqdockw{\ensuremath{\forall}} \coqdocvar{i},
\coqref{Assumptions.is}{\coqdocinductive{is}}
\coqref{doc.ro}{\coqdocconstructor{ro}}
(\coqref{doc.lookup}{\coqdocconstructor{lookup}} \coqdocvar{i})	 
& (1-lookup) & (\ref{fig:lookupdate})\coqdoceol\\
\coqdocindent{5pt}
\ensuremath{|} \coqdef{doc.is
  update}{is\_update}{\coqdocconstructor{is\_update}}:
\coqdockw{\ensuremath{\forall}} \coqdocvar{i},
\coqref{Assumptions.is}{\coqdocinductive{is}}
\coqref{doc.rw}{\coqdocconstructor{rw}}
(\coqref{doc.update}{\coqdocconstructor{update}} \coqdocvar{i})
& (2-update) & (\ref{fig:lookupdate})\coqdoceol\\
\coqdocindent{5pt}
\ensuremath{|} \coqdef{doc.is pure
  ro}{is\_pure\_ro}{\coqdocconstructor{is\_pure\_ro}}:
\coqdockw{\ensuremath{\forall}} \coqdocvar{X} \coqdocvar{Y} (\coqdocvar{f}:
\coqref{doc.term}{\coqdocinductive{term}} \coqdocvar{X} \coqdocvar{Y}),
\coqref{Assumptions.is}{\coqdocinductive{is}}
\coqref{doc.pure}{\coqdocconstructor{pure}} \coqdocvar{f}
\ensuremath{\rightarrow} \coqref{Assumptions.is}{\coqdocinductive{is}}
\coqref{doc.ro}{\coqdocconstructor{ro}} \coqdocvar{f}  
& (0-to-1) & (\ref{fig:meqn}) \coqdoceol\\
\coqdocindent{5pt}
\ensuremath{|} \coqdef{doc.is ro
  rw}{is\_ro\_rw}{\coqdocconstructor{is\_ro\_rw}}:
\coqdockw{\ensuremath{\forall}} \coqdocvar{X} \coqdocvar{Y}  (\coqdocvar{f}:
\coqref{doc.term}{\coqdocinductive{term}} \coqdocvar{X} \coqdocvar{Y}),
\coqref{Assumptions.is}{\coqdocinductive{is}}
\coqref{doc.ro}{\coqdocconstructor{ro}} \coqdocvar{f} \ensuremath{\rightarrow}
\coqref{Assumptions.is}{\coqdocinductive{is}}
\coqref{doc.rw}{\coqdocconstructor{rw}} \coqdocvar{f}
& (1-to-2) & (\ref{fig:meqn}) \coqdoceol\\
\end{tabular}
\coqdocemptyline
The decorated functions stated above are classified into four different manners: 
	\begin{itemize}
		\item terms specific to states effect: \coqdocconstructor{is\_lookup} and \coqdocconstructor{is\_update}
		\item categorical terms: \coqdocconstructor{is\_id}, \coqdocconstructor{is\_comp} and \coqdocconstructor{is\_final}
		\item terms related to categorical products: \coqdocconstructor{is\_pair}, \coqdocconstructor{is\_pi1} and \coqdocconstructor{is\_pi2}. 
		\item term decoration conversions based on the operation hierarchy:  \coqdocconstructor{is\_pure\_ro} and \coqdocconstructor{is\_ro\_rw}.  
	\end{itemize}
The \coqdocinductive{term} \coqdocconstructor{comp} enables one to compose two
compatible functions while 
the constructor \coqdocconstructor{is\_comp} enables one to compose functions
and to preserve their common decoration. 
For instance, if a \coqdocconstructor{ro} function is composed with another
\coqdocconstructor{ro}, then the composite function becomes
\coqdocconstructor{ro} as well.
For the case of the \coqdocconstructor{pair}, the same idea is used.
Indeed, the constructor \coqdocconstructor{is\_pair} takes two terms agreeing on
domains such as \coqdocinductive{term} \coqdocvar{$Y_1$ X}, say an
\coqdocconstructor{ro}, and \coqdocinductive{term} \coqdocvar{$Y_2$ X}, which is
\coqdocconstructor{ro} as well. 
\coqdocconstructor{is\_pair} then asserts that the pair of these terms is another
\coqdocconstructor{ro}. 
It is also possible to create both compositions and pairs of functions with
different decorations via the hierarchy rule stated among decoration types.
This hierarchy is build via the last two constructors,
\coqdocconstructor{is\_pure\_ro} and \coqdocconstructor{is\_rp\_rw}. 
The constructor \coqdocconstructor{is\_pure\_ro} indicates the fact that if a
term is \coqdocconstructor{pure}, then it can be seen as
\coqdocconstructor{ro}. Lastly \coqdocconstructor{is\_ro\_rw} states that if a
term is \coqdocconstructor{ro}, then it can be seen as \coqdocconstructor{rw} as
well. 

Note that the details of building pairs with different decorations can be found
in the derived pairs module (\texttt{Pairs.v} in the library).

The terms \coqdocconstructor{final}, \coqdocconstructor{pi1} and
\coqdocconstructor{pi2} are all \coqdocconstructor{pure} functions since they do
not manipulate the state.
\coqdocconstructor{final} forgets its input argument(s) and returns
nothing. Although this property could make one think that it generates a sort of
side-effect, this is actually not the case. 
Indeed, it is the only pure function whose co-domain is the terminal object
($\unit$) and it is therefore used to simulate the execution of a program:
successive, possibly incompatible, functions can then be composed with this
intermediate forgetfulness of results. 

The \coqdocconstructor{lookup} functions are decorated with the keyword $\acc$, as accessors. 
The constructor \coqdocconstructor{is\_lookup} is used to check the validity of
the \coqdocconstructor{lookup}\rq{} decoration. 
The different \coqdocconstructor{update} functions are \coqdocconstructor{rw}
and decorated with the keyword $\modi$. Similarly,  the constructor \coqdocconstructor{is\_update}
is thus used to check the validity of the \coqdocconstructor{update}\rq{}
decoration. 

		\subsection{Axioms}
		\label{sec:axioms}
We can now detail the Coq implementations of the axioms used in the proof
constructions. They use the given monadic equational logic and categorical
products.
The idea is to decorate also the equations. 
On the one hand, the \emph{weak} equality between parallel morphisms models the
fact that those morphisms return the same value but may perform different
manipulations of the state. 
On the other hand, if both the returned results and the state manipulations are identical, then the equality becomes \emph{strong}.

In order to define these decorations of equations in Coq, we again use inductive
terms and preserve the naming strategy of Section~\ref{sec:lefep}.

Below are given the reserved notations for \emph{strong} and \emph{weak} equalities, respectively.
\coqdocemptyline
\coqdocemptyline
\coqdocindent{3.0em}
\coqdocvar{Reserved Notation }\coqdocnotation{"} x \coqdocnotation{==} y\coqdocnotation{"} (\coqdoctac{at} \coqdocvar{level} 80). 
\coqdocindent{3.0em}
\coqdocvar{Reserved Notation }\coqdocnotation{"}x \coqdocnotation {$\sim$} y\coqdocnotation{"} (\coqdoctac{at} \coqdocvar{level} 80).
\coqdocemptyline
\coqdocemptyline

We have some number of rules stated w. r. t. \emph{strong} and \emph{weak}
equalities. The ones used in the proof given in Section \ref{sec:cupdlkp} are
detailed below. It is also worth to note that for each constructor of the given inductive types \coqdef{doc.strong}{strong}{\coqdocinductive{strong}} and \coqdef{doc.weak}{weak}{\coqdocinductive{weak}}, corresponding rules are shown in Appendix \ref{sec:decorules}, see Figures \ref{fig:meqn}, \ref{fig:empty-prod-existence} and \ref{fig:lookupdate}. 
\coqdocemptyline
\coqdocindent{3.0em}
\coqdockw{Inductive} \coqdef{doc.strong}{strong}{\coqdocinductive{strong}}: \coqdockw{\ensuremath{\forall}} \coqdocvar{X} \coqdocvar{Y}, \coqexternalref{relation}{http://coq.inria.fr/distrib/8.3pl4/stdlib/Coq.Relations.Relation\_Definitions}{\coqdocdefinition{relation}} (\coqref{doc.term}{\coqdocinductive{term}} \coqdocvar{X} \coqdocvar{Y}) :=
	\begin{itemize}
\item The rules \coqdef{strrefl}{strong\_refl}{\coqdocconstructor{strong\_refl}},  \coqdef{strsym}{strong\_sym}{\coqdocconstructor{strong\_sym}} and \coqdef{strtrans}{strong\_trans} {\coqdocconstructor{strong\_trans}} state that \emph{strong} equality is reflexive, symmetric and transitive, respectively. Obviously, it is an equivalence relation. See\rnsrefl,\rnssym and\rnstrans rules in Figure \ref{fig:meqn}.
\item Both \coqdef{idsrc}{id\_src}{\coqdocconstructor{id\_src}} and \coqdef{idtgt}{id\_tgt}{\coqdocconstructor{id\_tgt}} state that the composition of any arbitrarily selected function with \coqdocconstructor{id} is itself regardless of the composition order. See\rnidsrc and\rnidtgt rules in Figure \ref{fig:meqn}.
\item \coqdef{strsubs}{strong\_subs}{\coqdocconstructor{strong\_subs}} states that strong equality obeys the substitution rule. That means that for a pair of parallel functions that are \emph{strongly} equal, the compositions of the same source compatible function with those functions are still \emph{strongly} equal. \coqdef{strrepl}{strong\_repl}{\coqdocconstructor{strong\_repl}} states that for those parallel and  \emph{strongly} equal function pairs, their compositions with the same target compatible function are still \emph{strongly} equal.  See\rnssubs and \rnsrepl rules in Figure \ref{fig:meqn}.
\item
  \coqdef{rowts}{ro\_weak\_to\_strong}{\coqdocconstructor{ro\_weak\_to\_strong}}
  is the rule saying that all \emph{weakly} equal \coqdocconstructor{ro} terms are also \emph{strongly} equal. 
  Intuitively, from the given \emph{weak} equality, they must have the same
  results. Now, since they are not modifiers, they cannot modify the state. That means that effect equality requirement is also met. 
  Therefore, they are \emph{strongly} equal. See\rnws rule in Figure \ref{fig:meqn}.
\item \coqdef{cfu}{comp\_final\_unique} {\coqdocconstructor{comp\_final\_unique}} ensures that two parallel \coqdocconstructor{rw} functions (say \coqdocvar{f}
  and \coqdocvar{g}) are the same (\emph{strongly} equal) if they return the same
  result \texttt{f  $\sim$ g} together with the same effect
  \texttt{final $\circ$ f  == final $\circ$ g}. See Figure~\ref{fig:empty-prod-existence}.
	\end{itemize}
\coqdocindent{3.0em}
\coqdockw{with} \coqdef{doc.weak}{weak}{\coqdocinductive{weak}}:  \coqdockw{\ensuremath{\forall}} \coqdocvar{X} \coqdocvar{Y}, \coqexternalref{relation}{http://coq.inria.fr/distrib/8.3pl4/stdlib/Coq.Relations.Relation\_Definitions}{\coqdocdefinition{relation}} (\coqref{doc.term}{\coqdocinductive{term}} \coqdocvar{X} \coqdocvar{Y}) :=
	\begin{itemize}

\item \coqdef{zwrepl}{weak\_repl} {\coqdocconstructor{pure\_weak\_repl}} states that \emph{weak} equality obeys the substitution rule stating that for a pair of parallel functions that are \emph{weakly} equal, the compositions of those  functions with the same target compatible and \coqdocconstructor{pure} function are still \emph{weakly} equal. See\rnwrepl rule in Figure \ref{fig:meqn}.
\item \coqdef{stw}{strong\_to\_weak} {\coqdocconstructor{strong\_to\_weak}} states that \emph{strong} equality could be converted into
  \emph{weak} one, free of charge. Indeed, the definition of \emph{strong} equality
  encapsulates the one for \emph{weak} equality. See\rnsw rule in Figure \ref{fig:meqn}. 
\item \coqdef{ax2}{axiom\_2} {\coqdocconstructor{axiom\_2}} states that first updating a location \texttt{i} and then implementing an observation to another location \texttt{k} is \emph{weakly} equal to the operation which first forgets the value stored in the location \texttt{i} and observes location \texttt{k}. See\decatwo rule in Figure \ref{fig:lookupdate}. 
	\end{itemize}
Please note that \emph{weak} equality is an equivalence relation and obeys the substitution rule such as the \emph{strong} one. 

		\subsection{Derived Terms}
		\label{sec:ddwdec}
Additional to those explained in Section~\ref{sec:dndecterms}, some extra
terms are derived via the definitions of already
existing ones:
\coqdocemptyline
\coqdocindent{3.0em}
\coqdockw{Definition} \coqdef{doc.invpi1}{invpi1}{\coqd{inv\_pi1}} \{\coqdocvar{X} \coqdocvar{Y}\}: \coqref{doc.term}{\coqdocinductive{term}} (\coqdocvar{X}\coqexternalref{:type scope:x '*' x}{http://coq.inria.fr/distrib/8.3pl4/stdlib/Coq.Init.Datatypes}{\coqdocnotation{\ensuremath{\times}}}\coqdocvar{unit}) (\coqdocvar{X}) := \coqref{doc.pair}{\coqdocconstructor{pair}} \coqref{doc.id}{\coqdocconstructor{id}} \coqref{doc.unit}{\coqdocconstructor{unit}}.\coqdoceol
\coqdocindent{3.0em}
\coqdockw{Definition} \coqdef{doc.permut}{permut}{\coqd{permut}} \{\coqdocvar{X} \coqdocvar{Y}\}: \coqref{doc.term}{\coqdocinductive{term}} (\coqdocvar{X}\coqexternalref{:type scope:x '*' x}{http://coq.inria.fr/distrib/8.3pl4/stdlib/Coq.Init.Datatypes}{\coqdocnotation{\ensuremath{\times}}}\coqdocvar{Y}) (\coqdocvar{Y}\coqexternalref{:type scope:x '*' x}{http://coq.inria.fr/distrib/8.3pl4/stdlib/Coq.Init.Datatypes}{\coqdocnotation{\ensuremath{\times}}}\coqdocvar{X}) := \coqref{doc.Left pair}{\coqdocconstructor{pair}} \coqref{doc.pi2}{\coqdocconstructor{pi2}} \coqref{doc.pi1}{\coqdocconstructor{pi1}}.\coqdoceol
\coqdocindent{3.0em}
\coqdockw{Definition} \coqdef{doc.perm pair}{perm\_pair}{\coqd{perm\_pair}} \{\coqdocvar{X} \coqdocvar{Y} \coqdocvar{Z}\} (\coqdocvar{f}: \coqref{doc.term}{\coqdocinductive{term}} \coqdocvar{Y} \coqdocvar{X}) (\coqdocvar{g}: \coqref{doc.term}{\coqdocinductive{term}} \coqdocvar{Z} \coqdocvar{X}): \coqref{doc.term}{\coqdocinductive{term}} (\coqdocvar{Y}\coqexternalref{:type scope:x '*' x}{http://coq.inria.fr/distrib/8.3pl4/stdlib/Coq.Init.Datatypes}{\coqdocnotation{\ensuremath{\times}}}\coqdocvar{Z}) \coqdocvar{X}\coqdoceol
\coqdocindent{3.0em}
:= \coqref{doc.permut}{\coqd{permut}} \coqref{doc.::x 'o' x}{\coqdocnotation{o}} \coqref{doc.pair}{\coqdocconstructor{pair}} \coqdocvar{g} \coqdocvar{f}.\coqdoceol
\coqdocindent{3.0em}
\coqdockw{Definition} \coqdef{doc.prod}{prod}{\coqd{prod}} \{\coqdocvar{X} \coqdocvar{Y} \coqdocvar{X'} \coqdocvar{Y'}\} (\coqdocvar{f}: \coqref{doc.term}{\coqdocinductive{term}} \coqdocvar{X} \coqdocvar{X'}) (\coqdocvar{g}: \coqref{doc.term}{\coqdocinductive{term}} \coqdocvar{Y} \coqdocvar{Y'}): \coqref{doc.term}{\coqdocinductive{term}} (\coqdocvar{X}\coqexternalref{:type scope:x '*' x}{http://coq.inria.fr/distrib/8.3pl4/stdlib/Coq.Init.Datatypes}{\coqdocnotation{\ensuremath{\times}}}\coqdocvar{Y}) (\coqdocvar{X'}\coqexternalref{:type scope:x '*' x}{http://coq.inria.fr/distrib/8.3pl4/stdlib/Coq.Init.Datatypes}{\coqdocnotation{\ensuremath{\times}}}\coqdocvar{Y'})\coqdoceol
\coqdocindent{3.0em}
:= \coqref{doc.pair}{\coqdocconstructor{pair}} (\coqdocvar{f} \coqref{doc.::x 'o' x}{\coqdocnotation{o}} \coqref{doc.pi1}{\coqdocconstructor{pi1}}) (\coqdocvar{g} \coqref{doc.::x 'o' x}{\coqdocnotation{o}} \coqref{doc.pi2}{\coqdocconstructor{pi2}}).\coqdoceol
\coqdocindent{3.0em}
\coqdockw{Definition} \coqdef{doc.perm prod}{perm\_prod}{\coqd{perm\_prod}} \{\coqdocvar{X} \coqdocvar{Y} \coqdocvar{X'} \coqdocvar{Y'}\} (\coqdocvar{f}: \coqref{doc.term}{\coqdocinductive{term}} \coqdocvar{X} \coqdocvar{X'}) (\coqdocvar{g}: \coqref{doc.term}{\coqdocinductive{term}} \coqdocvar{Y} \coqdocvar{Y'}): \coqref{doc.term}{\coqdocinductive{term}} (\coqdocvar{X}\coqexternalref{:type scope:x '*' x}{http://coq.inria.fr/distrib/8.3pl4/stdlib/Coq.Init.Datatypes}{\coqdocnotation{\ensuremath{\times}}}\coqdocvar{Y}) (\coqdocvar{X'}\coqexternalref{:type scope:x '*' x}{http://coq.inria.fr/distrib/8.3pl4/stdlib/Coq.Init.Datatypes}{\coqdocnotation{\ensuremath{\times}}}\coqdocvar{Y'})\coqdoceol
\coqdocindent{3.0em}
:= \coqref{doc.perm pair}{\coqd{perm\_pair}} (\coqdocvar{f} \coqref{doc.::x 'o' x}{\coqdocnotation{o}} \coqref{doc.pi1}{\coqdocconstructor{pi1}}) (\coqdocvar{g} \coqref{doc.::x 'o' x}{\coqdocnotation{o}} \coqref{doc.pi2}{\coqdocconstructor{pi2}}).\coqdoceol
\coqdocemptyline

\texttt{Val\_i} and \texttt{Val\_i$\times\unit$} are isomorphic. Indeed, on the
one hand, let us form the left semi-pure pair $h=\rtuple{\id_{\Val_i},
  \tu_{\Val_i}}$. As $\tu_{\Val_i}$ is pure, then so is also $h$. Now, from the
definitions of semi-pure products 
the projections
yields $\pi_1\circ h \eqw \id_{\Val_i}$, which is also $\pi_1\circ h \eqs \id_{\Val_i}$ since
all the terms are pure. On the other hand, $\pi_1\circ
(h\circ\pi_1)\eqs\id_{\Val_i}\circ\pi_1\eqs\pi_1\eqs\pi_1\circ(\id_{\Val_i\times\unit})$
and
$\pi_2\circ(h\circ\pi_1)\eqs\tu_{\Val_i}\circ\pi_1\eqw\tu_{\Val_i\times\unit}\eqw\pi_2\eqs\pi_2\circ(\id_{\Val_i\times\unit})$.
but the latter weak equivalences are strong since all the terms are
pure. Therefore $\pi_1\circ(h\circ\pi_1)\eqs\pi_1\circ(\id_{\Val_i\times\unit})$
and $\pi_2\circ(h\circ\pi_1)\eqs\pi_2\circ(\id_{\Val_i\times\unit})$ so that
$h\circ\pi_1\eqs\id_{\Val_i\times\unit}$. Overall we have that $\pi_1$ is
invertible and $\pi_1^{-1}=h=\rtuple{\id_{\Val_i},\tu_{\Val_i}}$ as defined above.

We also have the \coqd{permut} term. It takes a product, switches the order of
arguments involved in the input product cone and returns the new product:
its signature is \coqdocinductive{term}
(\coqdocvar{Y}{\coqdocnotation{\ensuremath{\times}}}\coqdocvar{X})
(\coqdocvar{X}{\coqdocnotation{\ensuremath{\times}}}\coqdocvar{Y}).
 The \coqdocinductive{term} \coqd{perm\_pair} \coqdocvar{f} \coqdocvar{g} is handled via the composition of \coqdocconstructor{pair} \coqdocvar{g} \coqdocvar{f} with \coqd{permut}.
The definition \coqd{prod} is based on the definition of \coqdocconstructor{pair} with a difference that both input functions are taking a product object and returning another one while \coqd{perm\_prod} is the permuted version of \coqd{prod} which is built on \coqd{perm\_pair}s.

The decorations of \coqd{perm\_pair}, \coqd{prod} and \coqd{perm\_prod}, depend
on the decorations of their input arguments. For instance, a \coqd{perm\_pair}
of two \coqdocconstructor{pure} functions is also \coqdocconstructor{pure} while
the \coqd{prod} and \coqd{perm\_prod} of two \coqdocconstructor{rw}s is a
\coqdocconstructor{rw}.
These properties are provided by \coqd{is\_perm\_pair}, \coqd{is\_prod} and
\coqd{is\_perm\_prod}. More details can be found in the associated module of the
library (\texttt{Derived\_Terms.v}). 
Note that it is also possible to create \coqd{perm\_pair}s, \coqd{prod}s and
\coqd{perm\_prod}s of functions with different decorations via the hierarchy
rule stated among decoration types (\coqdocconstructor{is\_pure\_ro} and
\coqdocconstructor{is\_rp\_rw}). 
Existence proofs together with projection rules, can also be found in their
respective modules in the library (\texttt{Decorated\_Pairs.v} and \texttt{Decorated\_Products.v}).

		\subsection{Decorated Pairs}
		\label{sec:pair-exist-proj}
In this section we present some of the derived rules, related to \emph{pairs}
and \emph{projections}.
In Section~\ref{ssec:seqprod} we have defined the left semi-pure pair 
$\ltuple{\id_X,f}^\modi\colon X\to X\times Y$ of the identity $\id_X^\pure$
with a modifier $f^\modi\colon X\to Y$. 
In Coq this construction will be called simply the \verb!pair!  
of $\id_X^\pure$ and $f^\modi$.
The right semi-pure pair $\rtuple{f,\id_X}^\modi\colon X\to Y\times X$
of $f^\modi$ and $\id_X^\pure$ can be obtained 
as $\ltuple{\id_X,f}^\modi$ followed by the permutation $\perm_{X,Y}\colon
X\times  Y \to Y\times X$,
it will be called the \verb!perm_pair! of $f^\modi$ and $\id_X^\pure$. 

Then, the \coqref{doc.pair}{\coqdocconstructor{pair}} and \coqd{perm\_pair}
definitions, together with the hierarchy rules among function classes
(\coqref{doc.is pure ro}{\coqdocconstructor{is\_pure\_ro}} and  \coqref{doc.is
  ro rw}{\coqdocconstructor{is\_ro\_rw}}), are used to derive some other rules
related to existences and projections. 

\begin{itemize}
\item \coqdef{dpepurerw} {dec\_pair\_exists\_purerw}
  {\coqd{dec\_pair\_exists\_purerw}} is the rule that ensures that a
  \coqdocconstructor{pair} with a \coqdocconstructor{rw} and a
  \coqdocconstructor{pure} arguments also exists and is \coqdocconstructor{rw}
  too.
  \coqdef{wp1purerw} {weak\_proj\_pi1\_purerw} {\coqd{weak\_proj\_pi1\_purerw}}
  is the first projection rule stating that the first result of the pair is
  equal to the result of its first coefficient function. 
  In our terms it is given as follows: \texttt{pi1 $\circ$ pair f1 f2 $\sim$ f1}. The given equality is \emph{weak} since its left hand side is
  \coqdocconstructor{rw}, while its right hand side is \coqdocconstructor{pure}.
  \coqdef{sp2purerw} {strong\_proj\_pi2\_purerw}
  {\coqd{strong\_proj\_pi2\_purerw}} is the second projection rule of the
  semi-pure pair. 
  It states that the second result of the pair and its effect are equal to the
  result and effect of its second coefficient function. 
  In our terms it is given as follows: \texttt{pi2 $\circ$ pair f1 f2 == f2}.  \coqdef{dpermperwpure} {dec\_perm\_pair\_exists\_rwpure}
  {\coqd{dec\_perm\_pair\_exists\_rwpure}} is similar with pure and modifier
  inverted. 
\item \coqdef{dpepurero} {dec\_pair\_exists\_purero}
  {\coqd{dec\_pair\_exists\_purero}} is similar but with one
  coefficient function \coqdocconstructor{pure} and the other \coqdocconstructor{ro}. 
  Thus it must be an accessor by itself and its projections must be \emph{strongly} equal to
  its coefficient functions, since there is no modifiers involved. These properties are stated via 
  \coqdef{sp1purero} {strong\_proj\_pi1\_purero}
  {\coqd{strong\_proj\_pi1\_purero}} (\texttt{pi1 $\circ$ pair f1 f2 == f1})
  and \coqdef{sp2purero} {strong\_proj\_pi2\_purero}
  {\coqd{strong\_proj\_pi2\_purero}} (\texttt{pi2 $\circ$ pair f1 f2 ==
    f2}). \coqdef{dpermperopure} {dec\_perm\_pair\_exists\_ropure}
  {\coqd{dec\_perm\_pair\_exists\_ropure}} is similar with pure and accessor
  inverted. 
\end{itemize}
More details can be found in the \texttt{Decorated\_Pairs.v} source file.

		\subsection{Decorated Products}
		\label{sec:derprodexists}
Semi-pure products are actually specific types of semi-pure pairs,
as explained in Section~\ref{ssec:seqprod}. 
In the same way in Coq, the \verb!pair! and \verb!perm_pair! definitions 
give rise to the \verb!prod! and \verb!perm_prod! ones.

\begin{itemize}
\item   {\coqd{dec\_prod\_exists\_purerw}} ensures that a \coqd{prod} with a
  \coqdocconstructor{pure} and a \coqdocconstructor{rw} arguments exists and is \coqdocconstructor{rw}. 
  \coqdef{wpp1purerwr}{weak\_proj\_pi1\_purerw\_rect}
  {\coqd{weak\_proj\_pi1\_purerw\_rect}}  is the first projection rule
  and states that \texttt{pi1 $\circ$ (prod f g) $\sim$ f $\circ$ pi1}.
  {\coqd{strong\_proj\_pi2\_purerw\_rect}}  is the second projection rule and
  assures that \texttt{pi2 $\circ$ (prod f g) == f $\circ$ pi2}.
 Similarly, the rule \coqdef{dpermproderwpure}{dec\_perm\_prod\_exists\_rwpure}
  {\coqd {dec\_perm\_prod\_exists\_rwpure}} with projections: 
  \coqdef{spermp1rwpurer}{strong\_perm\_proj\_pi1\_rwpure\_rect}
  {\coqd{strong\_perm\_proj\_pi1\_rwpure\_rect}} (\texttt{pi1 $\circ$ (perm\_prod f g) == f $\circ$ pi1}) and \coqdef{wpermpp2rwpurer}{weak\_perm\_proj\_pi2\_rwpure\_rect}{\coqd{weak\_perm\_proj\_pi2\_rwpure\_rect}} (\texttt{pi2 $\circ$ (perm\_prod f g) $\sim$ g $\circ$ pi2}) relate to permuted products.

\item   {\coqd{dec\_prod\_exists\_purero}} ensures that a \coqd{prod} with a
  \coqdocconstructor{pure} and a \coqdocconstructor{ro} arguments exists and is \coqdocconstructor{ro}. 
  \coqdef{wpp1purerwr}{weak\_proj\_pi1\_purero\_rect}
  {\coqd{weak\_proj\_pi1\_purero\_rect}}  is the first projection rule
  and states that \texttt{pi1 $\circ$ (prod f g) == f $\circ$ pi1}.
  {\coqd{strong\_proj\_pi2\_purerw\_rect}}  is the second projection rule and
  assures that \texttt{pi2 $\circ$ (prod f g) == f $\circ$ pi2}. Similarly, permutation rule could be applied to get {\coqd{dec\_prod\_exists\_ropure}} rule with its projections: \coqdef{spermp1ropurer}{strong\_perm\_proj\_pi1\_ropure\_rect}
  {\coqd{strong\_perm\_proj\_pi1\_ropure\_rect}} (\texttt{pi1 $\circ$ (perm\_prod f g) == f $\circ$ pi1}) and \coqdef{wpermpp2ropurer}{weak\_perm\_proj\_pi2\_ropure\_rect}{\coqd{weak\_perm\_proj\_pi2\_ropure\_rect}} (\texttt{pi2 $\circ$ (perm\_prod f g) == g $\circ$ pi2}) 

\end{itemize}
For further explanation of each derivation with Coq
implementation, refer to the \texttt{Decorated\_Products.v} source file.
	\subsection{Derived Rules}
	\label{sec:derived}

The library also provides derived rules which can be just simple shortcuts for frequently used combinations of rules or more involved results. For instance: 

\begin{itemize}
	\item \coqdef{wrefl}{weak\_refl} {\coqd{weak\_refl}} describes the reflexivity property of the weak equality: It is derived from the reflexivity of the strong equality. 
\item Two pure functions having the same codomain $\unit$ must be strongly equal
  (no result and state unchanged). Therefore
  \coqdef{e03}{E\_0\_3}{\coqd{E\_0\_3}}  extends this to a composed function
  $f\circ g$, for two pure compatible functions \coqdocvar{f} and \coqdocvar{g},
  and another function \coqdocvar{h}, provided that \coqdocvar{g} and
  \coqdocvar{h} have $\unit$ as codomain. 
  \item In the same manner, \coqdef{e14}{E\_1\_4}{\coqd{E\_1\_4}} states that the composition of any
  \coqdocconstructor{ro} function \coqdocvar{h}$:\unit\to X$, 
  with \coqdocconstructor{final} is strongly equal to the
  \coqdocconstructor{id} function on $\unit$. 
  Indeed, both have no result and do not modify the state. 
\end{itemize}
More similar derived rules can be found in the \texttt{Derived\_Rules.v} source file.
	\section{Implementation of a proof: update-lookup commutation}
	\label{sec:cupdlkp}
We now have all the ingredients required to prove the update-lookup commutation 
property of Section~\ref{ssec:prop}

the order of operations
between updating a location \texttt{i} 
and retrieving the value at another location \texttt{j} does
not matter. 
The formal statement is given in Equation~(\ref{eq:cupdlkp}).

The value intended to be stored into the location \texttt{i} is an
element of \texttt{Val\_i} set while the lookup operation to the location
\texttt{j} takes nothing (apart from \texttt{j}), and returns a value read
from the set \texttt{Val\_j}. 
If the order of operations is reversed, then the element of \texttt{Val\_i} has
to be preserved while the other location is examined. Thus we need to form a
pair with the identity and create a product 
\texttt{Val\_i $\times \ \unit$}, via  \texttt{inv\_pi1}.
Similarly, the value recovered by the lookup operation has to be preserved and returned after the update operation. Then a pair with the identity is also created with update and a last projection is used to separate their results. The full Coq proof thus uses the following steps:

\begin{johnproof}
 \assumestep{\texttt{i, j:Loc}}
  \step{\texttt{lookup j $\circ$ update i == pi2 $\circ$ (perm\_prod (update i) id)}}{}
   \step{\texttt { $\circ$ (prod id (lookup j)) $\circ$ inv\_pi1}}{by \coqref{cfu}{\coqdocconstructor{comp\_final\_unique}}}
  \basestep[bs1]{1}
   \step{\texttt{final $\circ$ lookup j $\circ$ update i == final}}{}
   \step{\texttt{$\circ$ pi2 $\circ$ (perm\_prod (update i) id) }}{by \coqref{strsym}{ \coqdocconstructor{strong\_sym}}}
     \step{\texttt{$\circ$ (prod id (lookup j)) $\circ$ inv\_pi1}}{}
     \step{\texttt{final $\circ$ pi2 $\circ$ (perm\_prod (update i) id) }} {}
  \step{\texttt{$\circ$ (prod id (lookup j)) $\circ$ inv\_pi1}} {}
 \step{\texttt{ == final $\circ$ lookup j $\circ$ update i}}{by \coqref{strtrans}{\coqdocconstructor{strong\_trans}}}
 		\substep[ss11]{1.1}
 			 \step{\texttt{final $\circ$ pi2 $\circ$ (perm\_prod (update i) id)}}{}
			\step{\texttt{$\circ$ (prod id (lookup j)) $\circ$ inv\_pi1}} {}
			\step{\texttt{== pi1 $\circ$ (perm\_prod (update i) id) }}{}
   			\step{\texttt{$\circ$ (prod id (lookup j)) $\circ$ inv\_pi1}}{by  \coqref{e03}{\coqd{E\_0\_3}}}
\assumed
 		\substep[ss12]{1.2}
 			\step{\texttt{pi1 $\circ$ (perm\_prod (update i) id)}}{}
 			\step{\texttt{$\circ$ (prod id (lookup j)) $\circ$ inv\_pi1}}{}
 			\step{\texttt{== update i $\circ$ pi1 $\circ$ (prod id
                            (lookup j)) $\circ$ inv\_pi1}}{by \coqref{spermp1rwpurer}{\coqd{strong\_perm\_proj}}}
 \assumed	
 		\substep[ss13]{1.3}	
			\step{\texttt{update i $\circ$ pi1 $\circ$ (prod id (lookup j)) $\circ$ inv\_pi1}}{}
			\step{\texttt{== update i $\circ$ pi1 $\circ$ inv\_pi1}}{by \coqref{spp2purerwr}{\coqd{strong\_proj}}}
\assumed
 		\substep[ss14]{1.4}	
			\step{\texttt{update i $\circ$ pi1 $\circ$ inv\_pi1
                            == update i $\circ$ id}}{by \coqref{idtgt}{\coqdocconstructor{id\_tgt}} }
\assumed
 		\substep[ss15]{1.5}	
			\step{\texttt{final $\circ$ lookup j $\circ$ update i == update i $\circ$ id}}{by \coqref{e14}{\coqd{E\_1\_4}}}
\assumed \assumed
 \basestep[c1]{2}
     \step{\texttt{lookup j $\circ$ update i $\sim$ pi2 $\circ$ (perm\_prod (update i) id)}} {}
  \step{\texttt{$\circ$ (prod id (lookup j)) $\circ$ inv\_pi1}} {by \coqref{wtrans}{\coqdocconstructor{weak\_trans}}}

	  \substep[ss21]{2.1}
		\step{\texttt{lookup j $\circ$ update i $\sim$ lookup j $\circ$
                    final}} {by \coqref{ax2}{\coqdocconstructor{axiom\_2}}}
     \assumed  
	  \substep[ss22]{2.2}
		\step{\texttt{lookup j $\circ$ final $\sim$ lookup j $\circ$ pi2 $\circ$ inv\_pi1}}{see \S~\ref{sec:pair-exist-proj}}
     \assumed  
	  \substep[ss23]{2.3}
				\step{\texttt{pi2 $\circ$ (perm\_prod (update i) id)}}{}
				\step{\texttt{$\circ$ (prod id (lookup j)) $\circ$ inv\_pi1}}{}
				\step{\texttt{$\sim$ pi2 $\circ$ (prod id
                                    (lookup j)) $\circ$ inv\_pi1}}{by \coqref{spermp1rwpurer}{\coqd{strong\_perm\_proj}}}
     \assumed  
	  \substep[ss24]{2.4}
				\step{\texttt{pi2 $\circ$ (prod id (lookup j)) $\circ$ inv\_pi1}}{}
				\step{\texttt{$\sim$ lookup j $\circ$ pi2 $\circ$ inv\_pi1}}{by \coqref{spp2purerwr}{\coqd{strong\_proj}}}
     \assumed    
\assumed \assumed
  \step{\texttt{lookup j $\circ$ update i == pi2 $\circ$ (perm\_prod (update i) id)}}{}
   \step{\texttt {$\circ$ (prod id (lookup j)) $\circ$ inv\_pi1}}{}
\end{johnproof}
\coqdocemptyline
\coqdocemptyline
\coqdocindent{0.0em}
To prove such a proposition, the \coqref{cfu}{\coqdocconstructor{comp\_final\_unique}} rule
is applied first and results in two sub-goals to be proven: \texttt{final
  $\circ$ lookup j $\circ$ update i == final $\circ$ pi2 $\circ$
  (perm\_prod (update i) id) $\circ$ (prod id (lookup j)) $\circ$ inv\_pi1}
(to check if both hand sides have the same effect or not) and \texttt{lookup j
  $\circ$ update i $\sim$ pi2 $\circ$ (perm\_prod (update i) id) $\circ$ (prod
  id (lookup j)) $\circ$ inv\_pi1} (to see whether they return the same result
or not). Proofs of those sub-goals are given in step 1 and step 2, respectively.
\begin{enumerate}
  \renewcommand{\theenumi}{Step \arabic{enumi}}
\item \texttt{final $\circ$ lookup j $\circ$ update i == final
    $\circ$ pi2 $\circ$ (perm\_prod (update i) id) \\ $\circ$ (prod id (lookup
    j)) $\circ$ inv\_pi1}:
  \begin{enumerate}
    \renewcommand{\theenumii}{\arabic{enumi}.\arabic{enumii}}
  \item The left hand side 
     \texttt{final $\circ$ pi2
      $\circ$ (perm\_prod (update i) id) $\circ$ (prod id (lookup j)) $\circ$
      inv\_pi1}, (after the \coqref{strsym}{\coqdocconstructor{strong\_sym}} rule
    application) is reduced into: \texttt{pi1 $\circ$ (perm\_prod (update i) id)
      $\circ$ (prod (id (Val i)) (lookup j)) $\circ$ inv\_pi1}.
    The base point is an application of \coqref{e03}{\coqd{E\_0\_3}}, stating
    that \texttt{final $\circ$ pi2 == pi1} and followed by
    \coqref{strsubs}{\coqdocconstructor{strong\_subs}} applied to
    \texttt{(perm\_prod (update i) id)}, \texttt{(prod id (lookup j))} and
    \texttt{inv\_pi1}. 
  \item In the second sub-step,
    \coqref{spermp1rwpurer}{\coqd{strong\_perm\_proj\_pi1\_rwpure\_rect}} rule
    is applied to indicate the strong equality between \texttt{pi1 $\circ$
      perm\_prod (update i) id} and  \texttt{update i $\circ$ pi1}. After the
    applications of \coqref{strsubs}{\coqdocconstructor{strong\_subs}} with
    arguments \texttt{prod id (lookup j)} and \texttt{inv\_pi1}, we get:
    \texttt{pi1 $\circ$ (perm\_prod (update i) id) $\circ$ (prod id (lookup
      j)) $\circ$ inv\_pi1 == update i $\circ$ pi1 $\circ$ (prod id (lookup
      j)) $\circ$ inv\_pi1}. Therefore, the left hand side of the equation can
    now be stated as: \texttt{update i $\circ$ pi1 $\circ$ (prod id (lookup
      j)) $\circ$ inv\_pi1}.
  \item  Then, the third sub-step starts with the application of
    \coqref{wpp1purerwr}{\coqd{weak\_proj\_pi1\_purerw\_rect}} rule in order to
    express the following weak equality: \texttt{pi1 $\circ$ prod id (lookup
      j) $\sim$ id $\circ$ pi1}. 
    The next step is converting the existing weak equality into a strong one by
    the application of \coqref{rowts}{\coqdocconstructor{ro\_weak\_to\_strong}},
    since none of the components are modifiers. 
    Therefore we get: \texttt{pi1 $\circ$ prod id (lookup j) == id $\circ$
      pi1}. 
    Now, using \coqref{idtgt}{\coqdocconstructor{id\_tgt}}, we remove
    \texttt{id} from the right hand side. 
    The subsequent applications of
    \coqref{strsubs}{\coqdocconstructor{strong\_subs}} with arguments
    \texttt{inv\_pi1} and \coqref{strrepl}{\coqdocconstructor{strong\_repl}}
    enables us to relate \texttt{update i $\circ$ pi1 $\circ$ (prod id (lookup
      j)) $\circ$ inv\_pi1} with \texttt{update i $\circ$ pi1 $\circ$
      inv\_pi1} via a strong equality.
  \item In this sub-step, \texttt{update i $\circ$ pi1 $\circ$ inv\_pi1} is
    simplified into \texttt{update i $\circ$ id}. To do so, we start with
    \coqref{sp1purepure}{\coqd{strong\_proj\_pi1\_purepure}} so that
    \texttt{pi1 $\circ$ pair id final == id}, where \texttt{pair id
      final} defines \texttt{inv\_pi1=pair id final}. 
    Then, the application of \coqref{strrepl}{\coqdocconstructor{strong\_repl}}
    to \texttt{update i} provides: \texttt{update i $\circ$ pi1 $\circ$ pair
      id final == update i $\circ$ id}.
  \item In the last sub-step, the right hand side of the equation,
    \texttt{final $\circ$ lookup j $\circ$ update i}, is reduced into
    \texttt{update i $\circ$ id}.
    To do so, we use \coqd{E\_1\_4} which states that \texttt{final
      $\circ$ lookup j == id}. 
    Then, using \coqref{strsubs}{\coqdocconstructor{strong\_subs}} on
    \texttt{update i}, we get: \texttt{final $\circ$ lookup j $\circ$
      update i == id $\circ$ update i}. 
    By using \coqref{idtgt}{\coqdocconstructor{id\_tgt}} again we remove
    \texttt{id} on the right hand side and
    \coqref{idsrc}{\coqdocconstructor{id\_src}} rewrites \texttt{final
      $\circ$ lookup j $\circ$ update i} as \texttt{update i $\circ$ id}. 
  \end{enumerate}
  \seti
\end{enumerate}
At the end of the third step, the left hand side of the equation is reduced into
the following form: \texttt{update i $\circ$ id} via a \texttt{strong}
equality. Thus, in the fourth step, it was sufficient to show
\texttt{final $\circ$ lookup j $\circ$ update i == update i $\circ$ id} to
prove \texttt{final $\circ$ pi2 $\circ$ (perm\_prod (update i) id))
  $\circ$ (prod id (lookup j)) $\circ$ inv\_pi1 == final $\circ$ lookup
  j $\circ$ update i}. This shows that both sides have the same effect on the
state structure. 

\begin{enumerate}
  \renewcommand{\theenumi}{Step \arabic{enumi}}
  \conti
\item We now turn to the second step of the proof, namely: \texttt{lookup j
    $\circ$ update i $\sim$ pi2 $\circ$ (perm\_prod (update i) id) $\circ$
    (prod id (lookup j)) $\circ$ inv\_pi1}. The results returned by both input
  composed functions are examined. Indeed, from step 1 we know that they have
  the same effect and thus if they also return the same results, then they we
  will be strongly equivalent. 
  \begin{enumerate}
    \renewcommand{\theenumii}{\arabic{enumi}.\arabic{enumii}}
  \item Therefore, the first sub-step starts with the conversion of the left
    hand side of the equation, \texttt{lookup j $\circ$ update i}, into
    \texttt{lookup j $\circ$ final} via a weak equality. 
    This is done by the application of the
    \coqref{ax2}{\coqdocconstructor{axiom\_2}} stating that \texttt{lookup j
      $\circ$ update i $\sim$ lookup j $\circ$ final} for \texttt{j $\neq$
      i}.
  \item The second sub-step starts with the application of
    \coqref{sp2purepure}{\coqd{strong\_proj\_pi2\_purepure}} which states
    \texttt{pi2 $\circ$ (pair id final) == final} still with
    \texttt{(pair id final) = inv\_pi1}. 
    Then, via the applications of
    \coqref{strrepl}{\coqdocconstructor{strong\_repl}}, with argument
    \texttt{lookup j}, \coqref{stw}{\coqdocconstructor{strong\_to\_weak}} and
    \coqref{strsym}{\coqdocconstructor{strong\_sym}}, we get: \texttt{lookup j
      $\circ$ final $\sim$ lookup j $\circ$ pi2 $\circ$ inv\_pi1}. 
  \item  In the third sub-step, the right hand side of the equation,
    \texttt{pi2 $\circ$ (perm\_prod (update i) id) $\circ$ (prod id (lookup
      j)) $\circ$ inv\_pi1}, is simplified by \texttt{weak} equality: we start
    with the application of
    \coqref{wpermpp2rwpurer}{\coqd{weak\_perm\_proj\_pi2\_rwpure\_rect}} since
    \texttt{pi2 $\circ$ (perm\_prod (update i) id) $\sim$ id $\circ$
      pi2}. Then, we once again use
    \coqref{idtgt}{\coqdocconstructor{id\_tgt}} to remove the identity and the
    applications of \coqref{wsubs}{\coqdocconstructor{weak\_subs}} with
    arguments \texttt{prod id (lookup j)} and \texttt{inv\_pi1} yields the
    following equation: \texttt{pi2 $\circ$ (perm\_prod (update i) id) $\circ$
      (prod id (lookup j)) $\circ$ inv\_pi1 $\sim$ pi2 $\circ$ (prod id
      (lookup j)) $\circ$ inv\_pi1}.
  \item  In the last sub-step, \texttt{pi2 $\circ$(prod id (lookup j)) $\circ$
      inv\_pi1} is reduced into \texttt{lookup j $\circ$ pi2 $\circ$
      inv\_pi1} via a weak equality using
    \coqref{spp2purerwr}{\coqd{strong\_proj\_pi2\_purerw\_rect}} so that
    \texttt{pi2 $\circ$ (prod id (lookup j)) == lookup j $\circ$ pi2}. 
    Then, \coqref{strrepl}{\coqdocconstructor{strong\_repl}} is applied with
    argument \texttt{inv\_pi1}. 
    Finally, the \coqref{stw}{\coqdocconstructor{strong\_to\_weak}} rule is used
    to convert the \texttt{strong} equality into a \texttt{weak} one.
  \end{enumerate}
Both hand side operations return the same results so
that the statement \texttt{lookup j $\circ$ update i $\sim$ pi2 $\circ$
  (perm\_prod (update i) id) $\circ$ (prod id (lookup j)) $\circ$ inv\_pi1} is
proven.
\end{enumerate}	
Merging the two steps (same effect and same result) yields the proposition that
both sides are strongly equal. The full Coq development can be found in the
library in the source file \texttt{Proofs.v}. 
\qed

	\section{Conclusion}
	\label{sec:conclusion}

In this paper, we introduce a framework for the Coq proof assistant. The main
goal of this framework is to enable programmers to verify properties of programs
involving the global states effect.
We use a presentation of these properties which is close to syntax. In other
words, the state structure itself is not explicitly mentioned in the
verification progress. Instead, it is represented by the term decorations that
are used to declare program properties. 
We then used this framework to verify several well known properties of states as
the ones of~\cite{DBLP:conf/fossacs/PlotkinP02}. 
In order to verify these properties we first expressed them in the mathematical
environment of~\cite {DBLP:journals/corr/abs-1112-2396} where the effect of any
operation (function) is defined as the distance from being pure and is denoted
as \texttt{$\tu_Y \circ$ f} for any \texttt{f: X $\to$ Y}. 
Therefore, for the specific case of the global \texttt{state} effect, to check
for instance the \emph{strong} equality between any parallel morphisms
\texttt{f, g: X $\to$ Y}, we first check whether they have the same effect (this
is expressed via an equality \texttt{$\tu_Y \circ$ f $\eqs \tu_Y \circ$ g}) and
then monitor if they return the same result (this is expressed via a {\em weak}
equation \texttt{f $\eqw$ g}). This scheme has been integrally developed in Coq
and Section~\ref{sec:cupdlkp} illustrate the behavior of the resulting proofs on
one of the checked proofs of~\cite{DBLP:conf/fossacs/PlotkinP02}.

It is worth noting also that  the framework has been succesfully used to check a
more involved proof, namely that of Hilbert-Post completeness of the global
state effect in a decorated setting~\cite{Dumas:2013:patternsmonads}. The
process of writing this proof in our Coq
environment (now more than 16 Coq pages) for instance helped discovering at
least one non obvious flaw in a preliminary version of the proof.

Future work includes
extending this framework to deal with the {\em exception} effect: we know
  from~\cite{Dumas:2012:duality} that the {\em core} part of exceptions is dual
  to the global state effect. Then the extension would focus on the pattern
  matching of the handling of exceptions. 
We also plan to enable the verification of the {\em composition} of effects
and to extend the framework to other effects: for monadic or comonadic effects
  the generic patterns of~\cite{Dumas:2013:patternsmonads} could then be 
of help.

      \bibliographystyle{abbrv}
      \bibliography{myreferences}

\appendix

\section{Decorated rules for states}
\label{sec:decorules}

In order to prove properties of states, 
we introduce a set of rules which can be classified as follows: 
\emph{decorated monadic equational logic}, 
\emph{decorated categorical products} 
and \emph{observational products}. 
The full inference system can be found
in~\cite{DBLP:journals/corr/abs-1112-2396}. We give here a subset of
theses rules, enclosing those required for the proof of Section~\ref{sec:cupdlkp}.

\subsection{Rules for the decorated monadic equational logic}
\label{ssec:decorules-monadic}

From the usual categorical point of view, 
the rules of the \emph{monadic equational logic} are 
the rules for defining categories ``up to equations'': 
identities are terms, terms are closed under composition, 
the axioms for identities and associativity of composition 
are stated only up to equations, and the equations form a congruence.

For dealing with states, we use the \emph{decorated} version of 
the rules of the monadic equational logic
which is provided in Figure~\ref{fig:meqn}. 
These rules involve three kinds of terms ($\pure$, $\acc$ and $\modi$)
and two kinds of equations ($\eqs$ and $\eqw$); 
the meaning of these \emph{decorations} is given 
in Section~\ref{ssec:decorations}. 
The decoration $\dec$ stands for ``any decoration''. 

\begin{figure}[htb]
$$ \begin{array}{|c|} 
\hline 
\rnpid \dfrac{X}{\id_X^\pure:X\to X }  \qquad
\rncomp \dfrac{f^\dec:X\to Y \quad g^\dec:Y\to Z}{(g\circ f)^\dec:X\to Z} \quad
\rnpa \dfrac{f^\pure}{f^\acc} \quad
\rnpm \dfrac{f^\acc}{f^\modi} \\ \\
\rnsrefl \dfrac{f^{\modi}}{f \eqs f} \quad
\rnssym \dfrac{f^{\modi} \eqs g^{\modi}}{g \eqs f} \quad 
\rnstrans\dfrac{f^{\modi} \eqs g^{\modi} \quad g^{\modi} \eqs h^{\modi}}{f \eqs h} \\ \\
\rnassoc \dfrac{f^\modi:X\to Y \quad g^\modi:Y\to Z \quad h^\modi:Z\to W}
  {h\circ (g\circ f) \eqs (h\circ g) \circ f}  \quad
\rnidsrc \dfrac{f^{\modi}:X\to Y}{f\circ \id_X \eqs f} \quad 
\rnidtgt \dfrac{f^{\modi}:X\to Y}{\id_Y\circ f \eqs f} \\ \\
\rnssubs \dfrac{f^{\modi}:X\to Y \quad g_1^{\modi}\eqs g_2^{\modi}:Y\to Z}{g_1\circ f \eqs g_2\circ f :X\to Z}  \quad 
\rnsrepl \dfrac{f_1^{\modi} \eqs f_2^{\modi}:X\to Y \quad g^{\modi}:Y\to Z}{g\circ f_1 \eqs g \circ f_2:X\to Z} \\ \\
\rnws \dfrac{f^\acc \eqw g^\acc}{f \eqs g} \quad
\rnsw \dfrac{f^{\modi} \eqs g^{\modi}}{f \eqw g} \quad
\rnwsym \dfrac{f^{\modi} \eqw g^{\modi}}{g \eqw f} \quad 
\rnwtrans \dfrac{f^{\modi} \eqw g^{\modi} \quad g^{\modi} \eqw h^{\modi}}{f \eqw h} \\ \\ 
\rnwsubs \dfrac{f^{\modi}:X\to Y \quad g_1^{\modi}\eqw g_2^{\modi}:Y\to Z}{g_1\circ f \eqw g_2\circ f:X\to Z}  \qquad 
\rnwrepl \dfrac{f_1^{\modi}\eqw f_2^{\modi}:X\to Y \quad g^\pure:Y\to Z}{g \circ f_1 \eqw g\circ f_2:X\to Z} \\
\hline 
\end{array}$$
\caption{Rules of the decorated monadic equational logic for states}
\label{fig:meqn}
\end{figure}

For instance, the rule\rnpa says that if a function is pure, 
then it can be treated as an accessor, while the rule\rnpm 
says that an accessor can be treated as a modifier. 
The rule\rnwrepl says that the replacement rule for $\eqw$ 
holds for pure terms, but there is no general replacement rule for $\eqw$:
if $f_1^{\modi}\eqw f_2^{\modi}:X\to Y$ and $g^\acc:Y\to Z$ or $g^\modi:Y\to Z$,
then in general it cannot be proved that $g \circ f_1 \eqw g\circ f_2$: 
indeed, this property does not hold in the intended models.

\subsection{Rules for the decorated finite categorical products}
\label{ssec:decorules-prod}

The rules of the usual \emph{equational logic} are made of 
the rules of the monadic equational logic together with 
the rules for all finite categorical products ``up to equations'', 
or equivalently, 
the rules for a terminal object (or empty product)
and for binary products ``up to equations''.
When dealing with states, we use the \emph{decorated} version of 
these rules, as described in Figures~\ref{fig:empty-prod-existence}, 
\ref{fig:bin-prod-existence} and~\ref{fig:dec_lpair_unicity_existence}.

\begin{figure}[htb]
$$ \begin{array}{|c|} 
\hline 
\finale  \dfrac{}{\ \ \ \ \ \unit \ \ \ \ \ } \qquad  
\rnpfinal \dfrac{X}{\tu_X^\pure: X\to \unit} \qquad  
\rnwfinalun \dfrac{f^\modi, \ g^\modi:X\to \unit}{f \eqw g} \\  \\
\deccompuni \dfrac{f^\modi, \ g^\modi:X \to Y \quad \tu_{Y}^\pure \circ f^\modi \eqs \tu_{Y}^\pure \circ g^\modi \quad f^\modi \eqw g^\modi} {f \eqs g} \\ 
\hline 
\end{array}$$
\caption{Rules of the decorated empty product for states}
\label{fig:empty-prod-existence}
\end{figure}

One of the most important rules given in this context is\deccompuni 
in Figure~\ref{fig:empty-prod-existence}, which compares both 
the effects of two given parallel functions (\texttt{f} and \texttt{g}) 
and their results. 
If they have the same effect ($\tu\circ f \eqs \tu\circ g$) 
and the same result ($f\eqw g$), 
then the rule says that they are strongly equal ($f\eqs g$).

The rule\wuncty in Figure~\ref{fig:dec_lpair_unicity_existence} 
is another important rule for parallel functions ($f$ and $g$) 
returning a pair of results. It compares the first and the second 
result of both functions with respect to weak equality 
($\pi_1 \circ f \eqw \pi_1 \circ g$ stands for the first comparison 
and $\pi_2 \circ f \eqw \pi_2 \circ g$ for the second one);
if both weak equalities hold, then the rule says that
the functions $f$ and $g$ are weakly equal ($f\eqw g$).

\begin{figure}[htb]
$$ \begin{array}{|c|} 
\hline
\prodexists \dfrac{X_1 \quad X_2} {X_1 \times X_2}\quad  
\zprjone \dfrac{X_1 \quad X_2}{\pi_{X_1, X_2, 1}^\pure: X_1 \times X_2 \to X_1} \quad
\zprjtwo \dfrac{X_1 \quad X_2}{\pi_{X_1, X_2, 2}^\pure: X_1 \times X_2 \to X_2}  \\
\hline 
\end{array}$$
\caption{Rules of the decorated binary products for states: Existence}
\label{fig:bin-prod-existence}
\end{figure}

\begin{figure}[htb]
$$ \begin{array}{|c|} 
\hline  
\lldecpairexists \dfrac{f_1^\dec: X\to Y_{1} \quad f_2^\dec:X\to Y_{2}} {<f_1,  f_2>^\dec: X \to Y_1 \times Y_2}\\ \\ 
\lldecpairprjone \dfrac{f_1^\acc: X\to Y_{1} \quad f_2^\modi:X\to Y_{2}} {\pi_{Y_1, Y_2, 1} \circ <f_1, f_2> \ \eqw f_1} \quad
\lldecpairprjtwo \dfrac{f_1^\acc: X\to Y_{1} \quad f_2^\modi:X\to Y_{2}} {\pi_{Y_1, Y_2, 2} \circ <f_1, f_2> \ \eqs f_2} \\ \\
\wuncty \dfrac{f^\modi, g^\modi:X \to Y_1\times Y_2 \quad  \pi_{Y_1, Y_2, 1}^\pure \circ f^\modi \eqw \pi_{Y_1, Y_2, 1} \circ g^\modi \quad  \pi_{Y_1, Y_2, 2}^\pure \circ f^\modi \eqw \pi_{Y_1, Y_2, 2}^\pure \circ g^\modi } {f \eqw g}  \\ 
\hline 
\end{array}$$
\caption{Rules of the decorated pairs for states: Existence \& Unicity}
\label{fig:dec_lpair_unicity_existence}
\end{figure}

\subsection{Rules for the observational products}
\label{ssec:decorules-obs}

The rules in Figure~\ref{fig:lookupdate} are dedicated to the 
operations for dealing with states:
the $\lookup$ operations for observing the state and the 
$\update$ operations for modifying it. 
Let $\Loc$ denote the set of locations,
for each $i\in \Loc$ the type   
$V_i$ represents the set of possible values that can be stored 
in the location $i$, while $lookup_i$ and $update_i$ correspond to 
the basic operations that can be performed on this location.

\begin{figure}[htb]
$$ \begin{array}{|c|} 
\hline
\multicolumn{1}{|l|}{\mbox{ for each }\;i\in\Loc:}  \\
\dfrac {\ \ \ \ \ \ \ } {V_i}\quad 
\declkp \dfrac {\ \ \ \ \ \ \ } {\lookup_i^\acc: \unit \to V_i}\quad
\decupd \dfrac {\ \ \ \ \ \ \ } {\update_i^\modi: V_i \to \unit}\\ \\
\multicolumn{1}{|l|}{\mbox{ for each }\;i,k\in\Loc,\; i\ne k:}  \\
\decaone \dfrac {}{\lookup_i \circ \update_i \eqw \id_i} \quad
\decatwo \dfrac {}{\lookup_i \circ \update_k \eqw \lookup_i \circ \tu_k} \\ \\
 \\	
\decloctoglob \dfrac{f^{\modi}, g^{\modi}:X\to \unit \quad 
 \mbox{for each } \ k\in\Loc, \ \lookup_k^\acc \circ f^{\modi} \eqw \lookup_k^\acc \circ g^{\modi}} {f \eqs g} \\ 
\hline 
\end{array}$$
\caption{Rule of the decorated observational products for states}
\label{fig:lookupdate}
\end{figure}

The rule\decaone states that by updating a location $i$
and then reading the value that is stored in the same location $i$, 
one gets the input value. 
The equation is weak: indeed, 
the left hand side returns the same result as the right hand side 
but they have different state effects: 
$\lookup_i \circ \update_i $ is a modifier while $\id_i $ is pure. 

The rule\decatwo indicates that by updating a location $i$
and then reading the value that is stored in another location $k$, 
one gets the value stored in the location $i$. 
Besides, forgetting the value stored in the location $k$ 
and reading the one located in $i$, one gets as well the value stored in $i$. 
The equation is weak, since both hand side return the same result 
but they have different state effects: 
$\lookup_i \circ \update_k $ is a modifier while 
$\lookup_i \circ \tu_k$ is an accessor.

The rule\decloctoglob will be used for proving the strong equality of 
two parallel functions $f$ and $g$ (without result)  
by checking that the observed value at each location is the same
after modifying the state according to $f$ or according to $g$. 
Thus, many local observations yield a global result.

\end{document}